\documentclass[final,3p,12pt]{elsarticle}

\usepackage{graphicx}
\usepackage{ifpdf}
\usepackage{pdfpages} 
\usepackage{graphicx} 
\usepackage{adjustbox} 
\usepackage{color} 
\usepackage{enumerate} 
\usepackage{geometry} 
\usepackage{amsmath} 
\usepackage{amssymb} 
\usepackage[mathletters]{ucs} 
\usepackage[utf8x]{inputenc} 
\usepackage{fancyvrb} 
\usepackage{grffile} 
\usepackage{hyperref}
\usepackage{longtable} 
\usepackage{booktabs}  
\usepackage{float}
\definecolor{cream}{RGB}{222,217,201}
\usepackage{xcolor}
\usepackage{adjustbox} 
\newcommand{\prox}{\ensuremath{\operatorname{prox}}}
\newcommand{\sign}{\ensuremath{\operatorname{sign}}}

\newcommand{\ent}{\ensuremath{\operatorname{ent}}}
\newcommand\eC{\ensuremath{\mathbb{C}}}
\newcommand\eR{\ensuremath{\mathbb{R}}}

\newcommand\Hb{\ensuremath{\pmb{H}}}
\newcommand\Bb{\ensuremath{\pmb{B}}}
\newcommand\WL{\ensuremath{\mathcal{W}}}
\newcommand{\minimise}[2]{\ensuremath{\underset{\substack{{#1}}}%
{\mathrm{minimise}}\;\;#2 }}
\usepackage[]{natbib}

\begin{document}
\begin{frontmatter}



\title{PALMA, an improved algorithm for DOSY signal processing}


\author[LIGM,IGBMC]{A. Cherni}
\author[LIGM]{E. Chouzenoux}
\author[IGBMC]{M-A. Delsuc\corref{cor1}}
\ead{delsuc@igbmc.fr}
\address[LIGM]{Université Paris-Est, LIGM (UMR 8049), CNRS, ENPC, ESIEE Paris, UPEM, Marne-la-Vallée, France}
\address[IGBMC]{IGBMC, CNRS UMR 7104, 1 rue Laurent Fries BP10142, 67404 ILLKIRCH FRANCE}

\begin{abstract}
NMR is a tool of choice for the measure of diffusion coefficients of species in solution.
The DOSY experiment, a 2D implementation of this measure, has proven to be particularly useful for the study of complex mixtures, molecular interactions, polymers, etc.
However, DOSY data analysis requires to resort to inverse Laplace transform, in particular for polydisperse samples.
This is a known difficult numerical task, for which we present here a novel approach.
A new algorithm based on a splitting scheme and on the use of proximity operators is introduced.
Used in conjunction with a Maximum Entropy and $\ell_1$  hybrid regularisation, this algorithm converges rapidly and produces results robust against experimental noise.
This method has been called PALMA.
It is able to reproduce faithfully monodisperse as well as polydisperse systems, and numerous simulated and experimental examples are presented.
It has been implemented on the server \url{http://palma. labo.igbmc.fr} where users can have their datasets processed automatically.

\end{abstract}



\end{frontmatter}




\section{Introduction}

Diffusion coefficients can be efficiently measured in NMR by the use of magnetic field gradients.
The most classical approach consists in applying a symmetric pair of pulsed field gradients (PFG) of varying intensity, separated by a diffusion delay $\Delta$.
Random displacements of the molecule during $\Delta$ because of Brownian motion result in modulation of the signal intensity $I$ following the Stejskal-Tanner equation \cite{stejskal1965spin,Sinnaeve:2012jr}:
\begin{equation}
\label{eq:ST1}
I(q) = I_o exp( -D \Delta q^2 )
\end{equation}
where  $D$ is the diffusion coefficient of the molecular species and $q = \gamma \delta g$, the measure of the phase dispersion created by the PFG.
Here, $\gamma$ is the gyromagnetic ratio of the studied spin, and $\delta$ and $g$ are the duration and intensity of the PFG respectively.
A least squares fit of the experimental values to an exponential decay provides an estimate of the value of $D$.
The DOSY experiment, introduced by Johnson \cite{johnson1999diffusion} is a representation of this measure as a 2D spectrum, with chemical shifts presented horizontally and diffusion coefficients vertically.
DOSY has been used intensively to analyse molecular interactions, to sort the component in complex mixtures, or to evaluate molecular size distributions\cite{Morris:2007fy,Price:2009,Callaghan:2011}.

A monodisperse sample presents a well defined diffusion coefficient, and a simple exponential adjustment of equation (\ref{eq:ST1}) allows the determination of $D$.
When several compounds share the same chemical shift, resulting in overlapping lines in the NMR spectrum, the result of a mono-exponential fit becomes incorrect.
A simple column-wise least squares fit to two or more exponentials presents instabilities in noisy data-sets which make this approach difficult to use on complex cases.

Several methods have emerged in the literature for the analysis of complex mixtures of monodisperse species, where the difficult mostly arises from the presence in the spectrum of many overlapping species 
Approaches based on a global analysis of the whole experimental matrix have been proposed, based on a clever decomposition of the 2D spectrum matrix into multivariate model, which allows to extract the spectra of each species along with their respective diffusion profiles\cite{Stilbs:1996,Gorkom:1998}.
Some developments on this approach have been based on a harmonic analysis of the decay\cite{Windig:1997,Armstrong:2003,Huo:2004uj,Nilsson:2008,Stilbs:2010,Martini:2013ge}.
The exponential hypothesis can even be relaxed using methods related to blind source deconvolution\cite{Nuzillard:1998,Nuzillard:2005,Colbourne:2011,Toumi:2013iy}.

All these approaches model the sample as a mixture of species with a characteristic decay pattern.
Some samples such as polydisperse polymers, dendrimers, nanoparticles, gels and aggregated species,
because of the variation in size, length or aggregate state of the different molecules in the sample, present a distribution of diffusion coefficients rather than a single coefficient.
Moreover, the presence of common decay patterns at different chemical shifts cannot really be assumed anymore, as in these complex systems a subtle coupling between the chemical shift and the size usually broadens the spectral line, with each spectral channel sampling a slightly different subset of the species, so that each spectral channel has to be processed independently.

For these strongly polydisperse samples a precise determination of the diffusion distribution is of great analytical importance, it is however a difficult task.
Because of this difficulty, polydispersity is commonly measured by a polydispersity index (PDI) defined as the ratio of the mass average molar mass $M_w$ to its number average molar mass $M_n$ : PDI~$=~M_w / M_n$.
This quantity characterises the breadth of the distribution independently of the details of its shape, a PDI of $1.0$ indicates monodispersity.
PDI is commonly measured by size exclusion chromatography, (electron) microscopy, light or X-Ray scattering, or even NMR-DOSY\cite{vieville2011polydispersity}.

A polydisperse sample has to be analysed with a distribution $X(D)$ of diffusion coefficients and equation \eqref{eq:ST1} becomes:
\begin{equation}
     \label{eq:ST2}
    I(q) = \int_{D_{\min}}^{D_{\max}} X(D) exp( -D \Delta q^2 ) dD
\end{equation}

Determining the distribution $X(D)$ from $I(q)$  requires to solve the Laplace inversion of the $q^2$ dependency of $I(q)$.

The shape of the distribution $X$ can be modelled by a Gaussian function or by any other symmetric or asymmetric analytical shape, and the parameters for this shape fitted to the experimental data\cite{Williamson:2016ch}. 
This straight-forward approach is very sensitive to the choice of the shape, and will fail if it is not well adapted to the data, or if the distribution contains several isolated massifs, and it should be used with care.

In this work, we present a general approach that solves the Laplace inversion problem presented in equation \eqref{eq:ST2}.
A new algorithmic approach based on a splitting scheme and on the use of proximity operators is introduced.
Used in conjunction with Maximum Entropy and $\ell_1$ regularisations, the algorithm is stable against experimental noise, reproduces faithfully monodisperse as well as polydisperse situations, and converges rapidly.


\section{Theory}
\subsection{Problem description}
We assume that the diffusion experiment was performed over a series of $M$ values of $q$ (by varying $\delta$, $g$ or both) and measured as a series of intensities $y_m$ for a given chemical shift value.
The problem stated by equation \eqref{eq:ST2} can be discretised to be solved numerically:
$$
\label{eq:STn} 
y_m = \sum_{n=1}^N x_n exp( -D_n \Delta q_m^2 )
$$
with $D_n$ ranging from $D_{\min}$ to $D_{\max}$.
As this expression is linear in $x_n$, it can be rewritten as follows:
\begin{equation}
\label{eq:STmat}
Y = \Hb {X} 
\end{equation}
where $Y = \{ y_m, 1 \leq m \leq M \}$ is the experimental series,
$X=\{ x_n, 1 \leq n \leq N \}$ is a sampling of the distribution,
and $\Hb$ is an $M \times N$ matrix with entries $\Hb_{m,n} = exp( -D_n \Delta q_m^2 )$.
In this work, we call $X$ the Laplace spectrum of $Y$.
Determining $X$ from the knowledge of $\Hb$ and $Y$ is an \emph{ill-posed} problem as the experimental points are inevitably tainted with noise, and $\Hb$ is usually a non-square matrix, with $N>M$. 
A simple inversion does not provide a valid solution, and one has to resort to alternative approaches.
 
\subsection{Lagrangian formulation}
A general approach for solving equation \eqref{eq:STmat} is to generate a solution ${X}$ that solves the following regularised minimisation problem:
\begin{equation}
\label{eq:regul}
\minimise{X \in \mathbb{R}^N}\, \|\Hb X -Y\|^2 + \mu \Psi(X)
\end{equation}
The first term evaluates the distance between data and the reconstruction, while the second term is the regulariser, the Lagrangian coefficient $\mu > 0$ acting as a weight between the two.
The regularisation function $\Psi$ is aimed at selecting among all possible distributions, the most natural one, given the experimental evidences, using some \emph{a-priori} information on the problem.
It is usually built as a measure of the \emph{cost} of the reconstruction (in terms of energy, information, number of signals, etc\ldots \emph{see below})
and tends to favour an empty spectrum.
Depending on the expression chosen for $\Psi$, the problem can be solved by different approaches.
The CONTIN method \cite{provencher1982constrained} solves this problem for $\Psi(X) = \|\Gamma X\|^2$,
where $\Gamma$ is a matrix that contains prior assumptions about the data.
Classical choices are $\Gamma=Id$ which selects the solution with the least energy, 
or the first or second derivatives thus removing fluctuations not required for a faithful reconstruction.
It has known a great success since its introduction more than 30 years ago, however it suffers from slow convergence and over-smoothed solutions.
Choosing the opposite of the entropy as the regularisation function
($\Psi(X) = \sum (x_n / a) \log(x_n / a) $)
allows to produce the distribution with the least information in the sense of Shannon \cite{Nityananda82}.
This Maximum Entropy (MaxEnt) approach has shown to be of great efficiency and robustness for solving the DOSY problem \cite{Delsuc:1998} and has been widely used.
However, because of the strong curvature of the entropy function, the classical implementations\cite{Skilling:1984ti,Delsuc:1998}
 of this approach are known to present slow convergence rates.
Kazimierzuck \emph{et al} \cite{urbanczyk2013iterative} proposed recently to use $\Psi(X) = \sum|x_n| = \|X\|_1$ where $\Psi$ is the $\ell_1$ norm of $X$.
Their approach relies on previous works that have shown that this is equivalent to select the spectrum with the less non-null values.
The principal advantage of their approach is to rely on recent major advances in the field of convex minimisation and compressed sensing.
The algorithm ITAMeD they developed is based on the soft thresholding approach and allows a rapid convergence toward the solution.
Enforcing a minimum number of non-null values in $X$ is a good approach when the sample is a mixture of monodisperse compounds, and the Laplace spectrum a set of sharp lines, it is not appropriate however for the analysis of polydisperse samples that may present very large distributions.
\citet{urbanczyk2016monitoring} recently extended this work  to minimising the $\ell_p$ norm
($\|X\|_p = \left(\sum{|x_n|^p}\right)^\frac{1}{p}$) with $1 \leq p \leq 2$.
As expected, the authors show that adapting the $p$ parameter to the kind of data allows to reconstruct spectra of various widths.
However, the choice of $p$ is somewhat \textit{ad hoc} and has to be adapted to each situation, additionally the authors rely here on the IRLS \cite{Kazimierczuk:2011jy} algorithm that is slower than ITAMeD.
The TRAIn method \cite{xu2013trust} that has been proposed recently is not explicitly based on a regularisation approach, but rather on an early stopping strategy, in conjunction with a trust region algorithm. This method is claimed to be efficient on polydisperse distributions.
The early stopping approach creates a bias in the final solution which can be assimilated to an implicit regularisation albeit with no analytical definition, and this could be seen as a deficiency\cite{Idier:2008}.

\subsection{Constrained formulation}
The Lagrangian coefficient $\mu$ involved in equation \eqref{eq:regul} may be difficult to adjust in practice.
However, one has often some precise knowledge about the level of noise corrupting the data.
We propose to adopt a more practical formulation by solving the following constrained optimisation problem:
\begin{equation}
\label{eq:pbc}
\minimise{X \in \eR^N}{\Psi(X)} \quad \text{subject to}\quad  \| \Hb X - Y \| \le \eta
\end{equation}
where $\eta > 0$ is related to the expected quality of the fit, based on an estimate of the experimental noise.
This has the advantage to shift the burden of determining the adequate value of a Lagrangian coefficient to the much simpler task of estimating a noise level.

\subsection{Hybrid regularisation}
In order to favour both smooth (polydisperse) and sparse (monodisperse) shapes in the estimated signal, we propose a novel regularisation defined as follows:
\begin{equation}
\label{eq:psi}
\Psi(X) =    \lambda \ent(X,a)  + (1 - \lambda) \ell_1(X)
\end{equation}
where $\ent{(X,a)}$ is given by
$$ 
\ent(X,a) =   
\begin{cases}
    \sum_{n=1}^N  \frac{x_n}{a} \log  \left(\frac{x_n}{a}\right) & \mbox{if} \, \,x_n > 0 \\
    0 & \mbox{if} \, \,  x_n = 0\\
    +\infty & \mbox{elsewhere},
\end{cases}
$$
and is the opposite of the Shannon entropy with a flat prior $a>0$, typically chosen here from an estimate value of $\sum x_n$.
$\ell_{1}(X)$ is the $\ell_1$ norm of the vector $X$ defined as \cite{eladCS}:
$$ 
\ell_{1}(X) = \sum_{n=1}^N  |x_n|
$$
and $\lambda \in [0,1]$ allows to control the balance between the sparsity prior and the entropy prior.  

\subsection{Proximity operator}
The general problem expressed in equation \eqref{eq:pbc} can be elegantly solved by the convex optimisation algorithm PPXA+, based on the use of proximity operators \cite{bauschke2011convex}.
This allows a generic solver for any choice of convex function $\Psi$ while insuring a very rapid convergence toward the optimal solution.
The proximal approach is also the basis of the ITAMeD method.
The FISTA algorithm on which it is based \cite{Beck09} is derived from the soft-thresholding operator, the proximity operator of the $\ell_1$ norm operator. 
To our knowledge, it has never been applied to the MaxEnt penalty nor to a hybrid approach as proposed here. 

It can be shown\cite{CherniISIVC2016} that the proximity operator of the functional $\Psi$ used in equation \eqref{eq:psi} can be expressed as follows (see ESI sections 1 and 2\dag\ for details):
\begin{equation}
\label{eq:prox0}
\prox_{ \Psi}(X) = (p(x_n))_{1 \leq n \leq N}
\end{equation}
where,  
\begin{equation} 
\label{eq:prox}
p(x_n) = \begin{cases}
    \frac{\lambda}{a}\WL
    \left[\frac{a}{\lambda}  \exp({\frac{a x_n - a (1 - \lambda)}{\lambda} + 
    \log(a) - 1})\right] 
    &  \mbox{if} \quad \lambda \in ]0,1]\\
    \sign(x_n) \max\left(\left|x_n\right| - (1 - \lambda), 0 \right)
    & \mbox{if} \quad \lambda = 0
\end{cases}     
\end{equation}
In the above expression, $\WL$ states for the Lambert function, defined as the inverse function of $f: z \rightarrow ze^z $ for all $z \in \eC$, \cite{corless1996lambertw} i.e:
$$
z = we^w   \quad \Leftrightarrow    \quad w = \WL(z).
$$
In the current context, only a restriction of $\WL$ to $\eR^{+}$ is required.
It should be noted that for pure Maximum Entropy $(\lambda = 1)$ we recover the result from \citet{combettes2011proximal}:
$$
p_{\ent}(x)= \frac 1 a \WL \left( a \exp(x) + \log(a) - 1 \right)
$$
Similarly, pure $\ell_1$ regularisation $(\lambda = 0)$ brings the soft thresholding operator:
$$
 p_{\ell_1}(x) =  \sign(x) \max\left(|x| - 1 , 0 \right)
$$

\subsection{Algorithm}
With the expression of the proximity operator given in equations \eqref{eq:prox0}-\eqref{eq:prox}, the convex optimisation problem \eqref{eq:pbc} can be easily solved using a proximal splitting algorithm.
At each iteration, such a method alternates between the proximity operator of $\Psi$, and the proximity operator associated to the constraint $\| \Hb X - Y \| \leq \eta,$
(i.e. the projection operator onto this constrained set).
 
In order to ensure good convergence properties of our algorithm, we adopt the PPXA+ approach from \citet{Pustelnik_2011_hybrid_regularization_restoration}, generalizing the PPXA method from \citet{Combettes2008PPXA}. These algorithms both rely on the Douglas-Rachford scheme \cite{Combettes2007DouglasRachford}, which consists in replacing the involved proximity operators by their reflections (see ESI sections 3, 4 and 5\dag\ for details).

This leads us to the so-called PALMA algorithm, standing for ``\textbf{P}roximal \textbf{A}lgorithm for \textbf{L}$_1$ combined with \textbf{MA}xent prior''.
This algorithm is fully detailed in the Electronic Supplementary Information\dag .

\section{Material \& Methods}

\subsection{Simulations.}
Several simulated data-sets, chosen to represent various analytical situations, were used for the evaluation of the algorithm.
Set A consists in three monodisperse components with diffusion coefficients $16\,\mu m^2/s$ , $63\, \mu m^2/s$ and $230\,\mu m^2/s$, with respective intensities $1.0$, $0.33$, $0.66$.
This data-set is equivalent to the simulation used in \citet{urbanczyk2013iterative}.
Set B is a wide distribution, simulated as a log-normal distribution centred at $35\,\mu m^2/s$,
it presents a PDI estimated to $6.26$.
Sets C1 and C2 are asymmetric distributions built from $15$ log-normal components, ranging from $18$ to $85$ $\mu m^2/s$, with intensities ranging from $0.1$ to $10$, they have PDI estimated respectively to $1.79$ and $1.32$.
In all simulations, $64$ gradient values were simulated, and a Gaussian noise equal to $0.001\%$, $0.01\%$, $0.1\%$, or $1\%$ of the initial point was added.
The gradient values were chosen with a harmonic progression for set A, and with linear increments for sets B, C1, and C2.
All Laplace spectra were reconstructed on 256 logarithmically sampled points.
Other simulations with varying distributions and noise levels are also presented in ESI (see section 6 \dag).

\subsection{NMR measure.}
A set of PEO standards were purchased from American Polymer Standards Corporation (Mentor, OH, USA),
and 3 samples were prepared.
Sample a) is a standard PEO with $M_w=2343.3 \, g\,mol^{-1}$ and PDI = $1.07$;
sample b) is a standard PEO with $M_w=4051.2 \, g\,mol^{-1}$ and PDI = $1.28$;
sample c) is a mixture prepared from standard PEOs ranging from $350$ to $5250\,g\,mol^{-1}$  for a theoretical $M_w$ of $3238.5\,g \,mol^{-1}$ and a theoretical PDI of $2.01$.
Each sample was prepared and measured as described in \citet{vieville2011polydispersity}.

The crude plant extract was obtained from the brown algae \textit{Sargassum muticum} as described in \citet{Vonthron:2011}.
Two equivalent samples were prepared by dissolving each time $18.3\,mg$ in $0.75\,mL$ MeOD plus $0.4\,mL$ D$_2$O,
chloroquine was added to one sample at a concentration of $0.16\,mg \,mL^{-1}$ ($1\% w/w$ of plant dry extract).
The NMR experiments were run on a Avance III Bruker spectrometer, operating at $700$\,MHz, and equipped with a TXI cryo-probe.
DOSY were acquired with the convection compensated experiment using bipolar pulses \cite{Jerschow:1997vm} (\texttt{dstebpgp3spr} pulse program).
$50$ gradient increments from $0.5\,G/cm$ to $52.5\,G/cm$ were used, with a cosine roll-off PFG shape.
Each elementary PFG had a duration of $\delta = 1.1\,ms$ and the diffusion delay $\Delta$ was set to $150\,ms$.
For each gradient intensity, $64$ scans were acquired with a relaxation of $1.5\,s$,
for a total experimental time of $1$\,hour $50$\,minutes.
Each 1D spectrum was apodised with an unshifted sine-bell, zerofilled once and Fourier transformed.
A spline baseline correction was applied, as well as a correction of small shifts caused by possible instabilities of the temperature control\cite{assemat:2010ti}.

\subsection{Processing.}
The PALMA algorithm was implemented with the programming language python version 2.7 using the numpy/scipy libraries.
The algorithm was then packaged in a plugin to the SPIKE program developed in our group\cite{spike}.
All programs are available from the authors.
All computations were performed on a Macintosh Mac Pro dual Xeon with a total of 8 cores, equipped with 32 GB of memory and running MacOsX 10.7. 

The DOSY experiments presented in Fig.\ref{fig:Poly} and Fig.\ref{fig:Chlo} were processed column-wise with $\lambda=0.01$.
The noise level along the gradient axis was estimated by the difference of the actual data to the result of a polynomial smoothing, this value being used to estimate $\eta$ in equation \eqref{eq:pbc} (See ESI, section 5.2\dag\ for more details).
The diffusion axis was logarithmically sampled from $50$ to $10.000\,\mu m^2 s^{-1}$ over $256$ points.

Experiments on PEO were processed with a maximum of $20.000$ iterations, for a total time of about $30$ seconds for $110$ DOSY profiles computed.
Experiments on plant extract were processed taking all signals with an estimated SNR above $20$ ($26\,dB$) in the first 1D spectrum of the experimental matrix, with a maximum of $200.000$ iterations, for a total time of about $1$ hour for $1200$ to $1400$ DOSY profiles.

The algorithms ITAMeD, $\ell_p$ tailored-ITAMeD, and TRAIn utilised in the ESI were used as downloaded from the respective web-sites, using MATLAB program version R\_2013b on MacOs.

\subsection{Server.}
A Web server is available at \url{http://palma.labo.igbmc.fr}, where users may submit data-sets for automatic processing.
The python code of the PALMA algorithm is available on the same server as well as on
\url{https://github.com/delsuc/PALMA}.

\section{Results}
\subsection{Tests on Simulated Data}

\begin{figure}[h!]
\centering
\includegraphics[width = 0.9\textwidth]{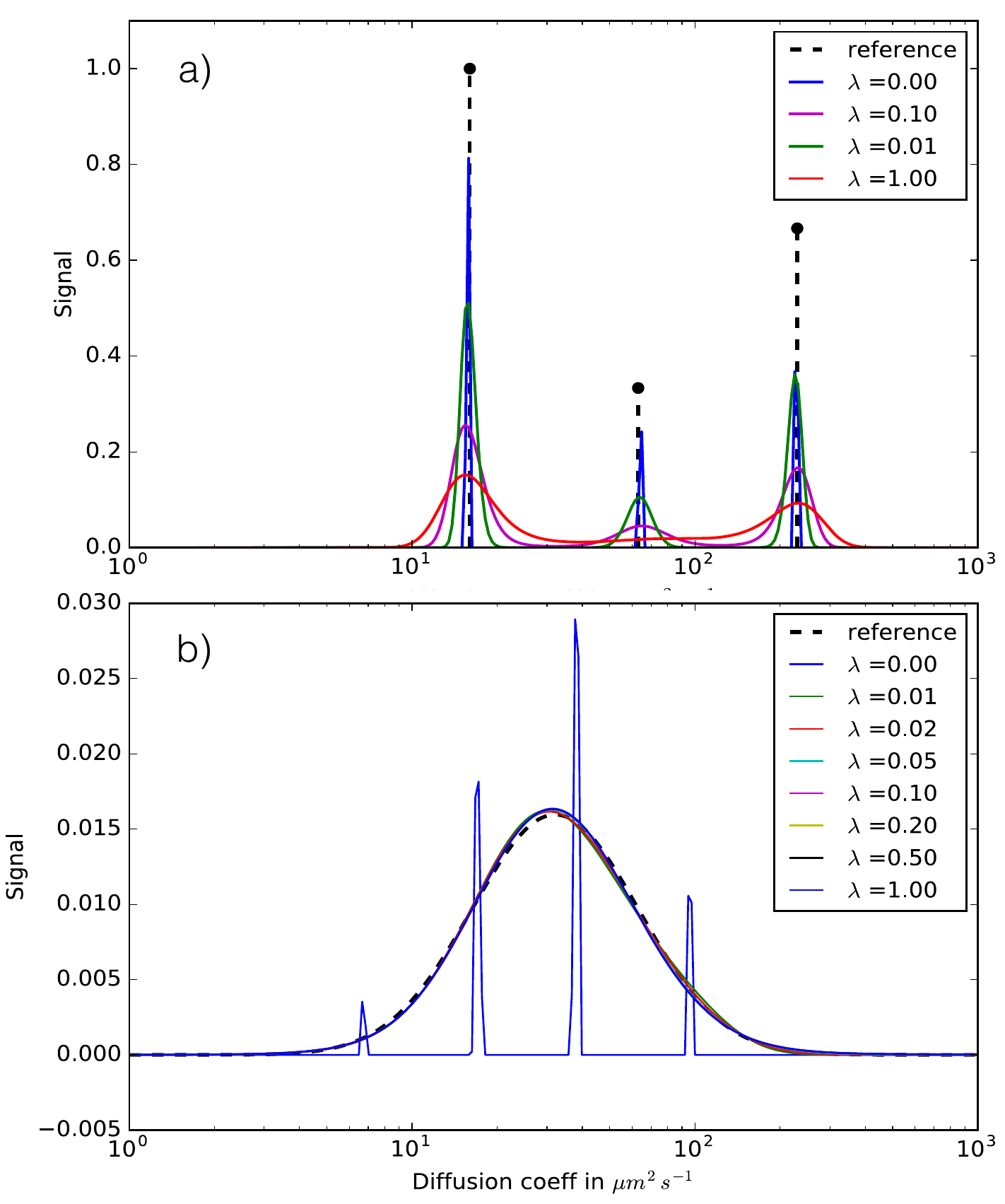}
\caption{PALMA Reconstruction for various values of $\lambda$ of the simulated experiments, with an added 0.1\% Gaussian noise.
a)  experiment A  with 3 monodisperse species, indicated by the black vertical lines, $\lambda=0$ spectrum was divided by $3$ for clarity;
b)  experiment B with a large polydisperse Gaussian profile $\lambda=0$ spectrum was divided by $8$ for clarity. }
\label{fig:Simu} 
\end{figure} 

The PALMA algorithm described above was first tested on a series of simulated data-sets.
Fig.\ref{fig:Simu}a presents the results obtained on the simulated experiment A consisting in the superposition of three monodisperse species, separated by less than a factor of $4$ in diffusion coefficients, equivalent to the test used in \citet{urbanczyk2013iterative}.
When analysed with PALMA using a null $\lambda$ value, indicating a pure $\ell_1$ regularisation, a Laplace spectrum consisting of 3 sharp peaks is produced as expected.
Using the pure MaxEnt mode ($\lambda=1$) on the same dataset, a broader spectrum is reconstructed.
PALMA allows the weight between the two approaches to be freely varied.
When doing so it can be observed a narrowing of the MaxEnt distribution for decreasing value of $\lambda$, characteristic of a bias toward monodisperse distributions, with a sudden transition to sharp lines for $\lambda=0$.
In Fig.\ref{fig:Simu}b the same procedure was applied on a broad Gaussian line simulating a polydisperse polymer with a PDI of about $6$, corresponding to experiment B.
Again it can be observed that a null $\lambda$ gives rise to sharp, sparse lines, this time only sampling the broad line in an inadequate manner.
However, in contrast with the previous case, all the non-null values of $\lambda$ lead to a nearly perfect reconstruction of the line-shape, with a correct determination of its width. In this example, the minimal reconstruction error was obtained for $\lambda = 0.05$.
Tests performed on asymmetric distributions (see  ESI figures S5 and S6 and table S1\dag ) show the same tendency, with stable results for all non-null values of $\lambda$.
It should be noted that, because of the algorithm that maintains the analysis within the noise distance of the data, i.e.  $ \| \Hb X - Y \| \lessapprox \eta$ (see \eqref{eq:pbc}) all the reconstructed Laplace spectra fit equally well the data.
They differ only in how they match the regularisation term, a term which holds and expresses the \emph{a priori} information we have on the data-set.

The PALMA algorithm was tested against ITAMeD\cite{urbanczyk2013iterative}, $\ell_p$ tailored-ITAMeD\cite{urbanczyk2016monitoring}, and TRAIn\cite{xu2013trust} algorithms, using the same simulated data as above.
Table \ref{table:1} presents synthetic results, extensive results are presented in ESI (see figures S7 to S14 and tables S2 and S3\dag ).

\begin{table}[H]
\centering 
\renewcommand{\arraystretch}{1.3} 
\begin{tabular}{cllll}\hline
 {Algorithm} & \multicolumn{4}{c}{noise level}  \\ 
 & $1 \%$ & $0.1 \%$  & $0.01 \%$  & $0.001 \%$ \\  \hline
 ITAMeD  & $3.37$  & $18.65$    & $29.04$    & $29.40$ \\ 
  ITAMeD with $\ell_p$   & $6.06$  & $25.26$    & $36.69$    & $37.08$ \\ 
  TRAIn           &  \textcolor{red}{$24.75$}  & $28.63$  & $26.53$   & $19.47$ \\ 
  PALMA  $\lambda=0.01$ &  $20.54$ & $28.57$   & $41.69$   & \textcolor{red}{$53.25$} \\ 
  PALMA  $\lambda=0.05$ &  $24.01$ & \textcolor{red}{$32.51$}   & \textcolor{red}{$48.28$}   & $51.37$ \\ \hline 
  \end{tabular} 
\caption{Quality of reconstruction of signal B with different algorithms for various noise levels.
Quality is computed as $\frac{\|X_{sim}\|}{\|X_{sim}-X_{calc}\|}$ expressed in dB.
For each noise level, the highest quality results are outlined.}
\label{table:1} 
\end{table}

In our hands, PALMA and TRAIn present the best results in terms of faithfulness and robustness, in particular for polydisperse data-sets, with TRAIN showing better results on the C2 data-set, while PALMA behaving better for data sampled with a small number of data-points.

\subsection{Application to polydisperse polymers.}
\begin{figure}[h!]
\centering
\includegraphics[width = 0.9\textwidth]{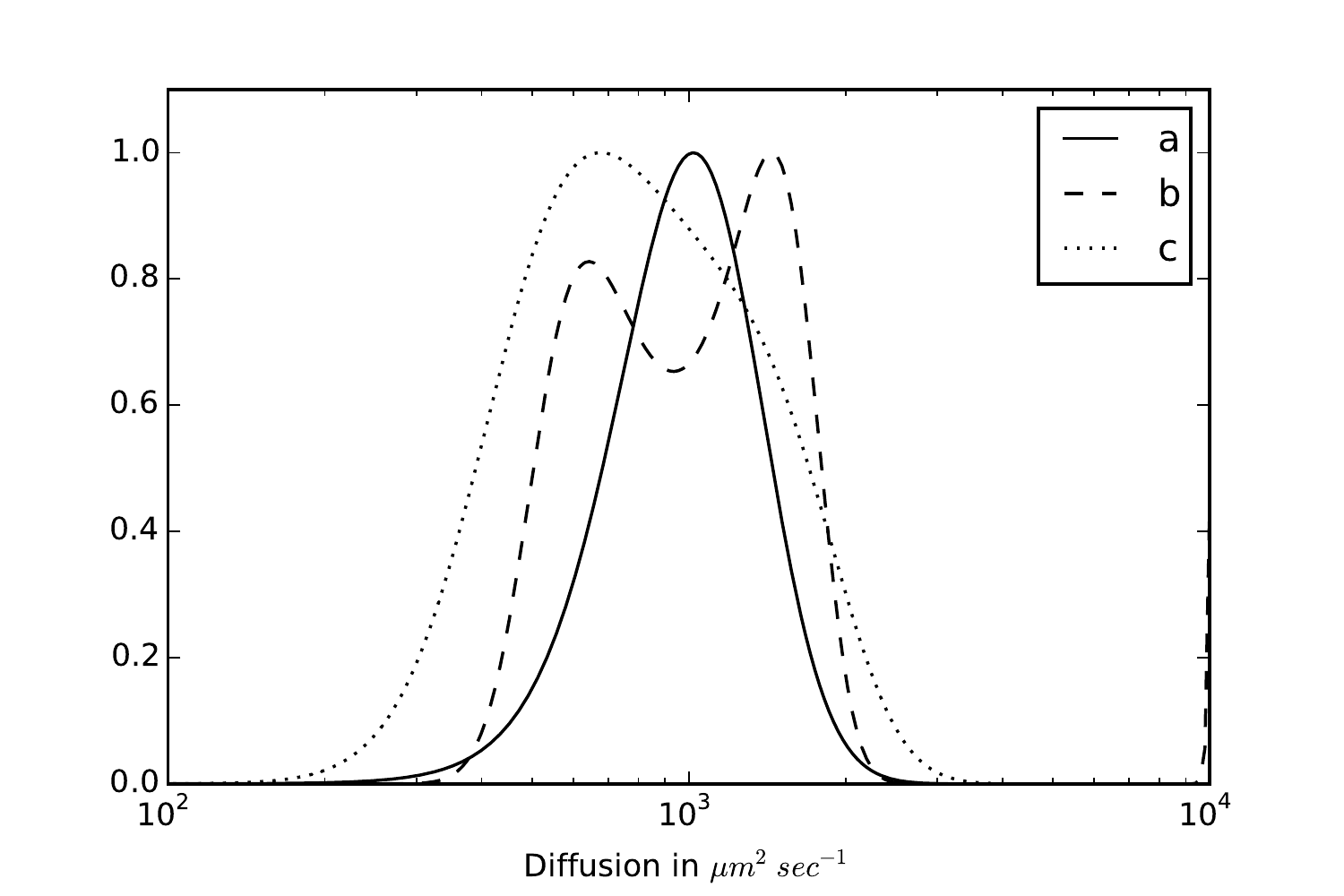}
\caption{DOSY profile of the main NMR signal of different standard PEO samples,
a) a reference PEO with PDI of $1.07$,
b) a reference PEO with PDI of $1.28$,
c) a mixtures of reference PEOs with a global PDI of $2.01$,
}
\label{fig:Poly}
\end{figure}

It is acknowledged that experimental data are quite different from simulated one, with a mixture of sharp and large diffusion distribution, tainted with instrumental artifacts and non-stationary noise.
To the behaviour of the method on polydisperse systems, the program was first applied on DOSY experiments measured from  poly-ethyleneoxide (PEO) polymers in water with calibrated chain lengths and polydispersity.
Fig. \ref{fig:Poly} presents the results obtained for three PEO samples with polydispersity ranging from $1.07$ to $2.0$ measured in a standard manner, and processed with PALMA.
The polydispersity of the different samples can be clearly seen in the profile widthes.
Sample a) is a standard PEO polymer, with a rather low polydispersity.
Sample c) was  prepared from a set of rather monodisperse polymers, in order to cover regularly a large range of chain lengths. 
Sample b) on the other hand is a  PEO polymer given to have a standard polydispersity, however the details of the composition is not known.

\subsection{Application to plant extract.}
To test the robustness of the approach, it was applied to crude ethanolic plant extract obtained from brown algae.

\begin{figure}[h!]
\centering
\includegraphics[width = 0.9\textwidth]{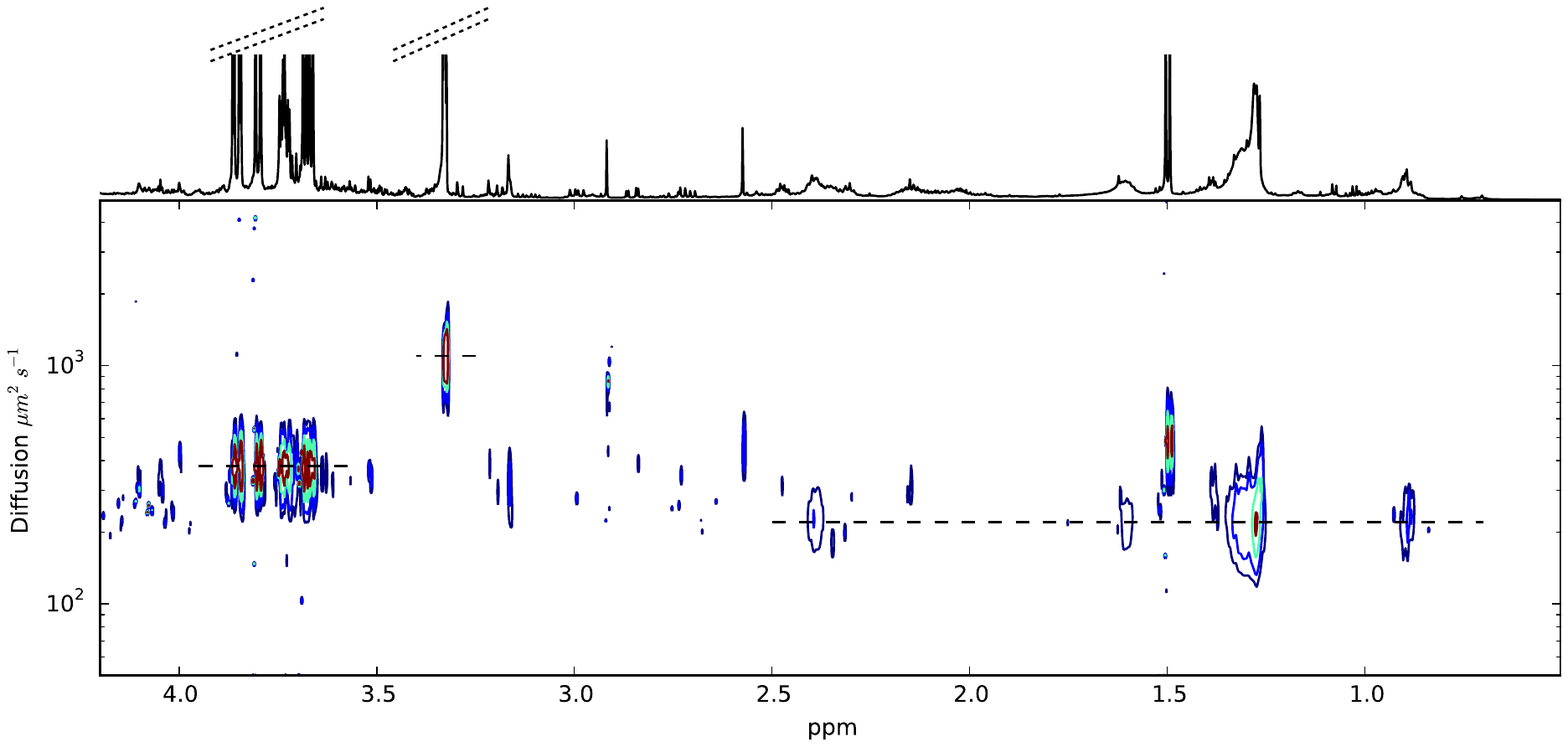}
\caption{DOSY experiment on a brown algae methanol/water extract showing only the major constituents.
Dashed horizontal label lines are indicated for
fatty acid chains ($220\, \mu m^2 s^{-1}$), glycerol and short polyol ($380 \, \mu m^2 s^{-1}$), methanol ($1100 \, \mu m^2 s^{-1}$).
} 
\label{fig:Chlo}
\end{figure}

\begin{figure*}[h!]
\includegraphics[width = 0.98\textwidth]{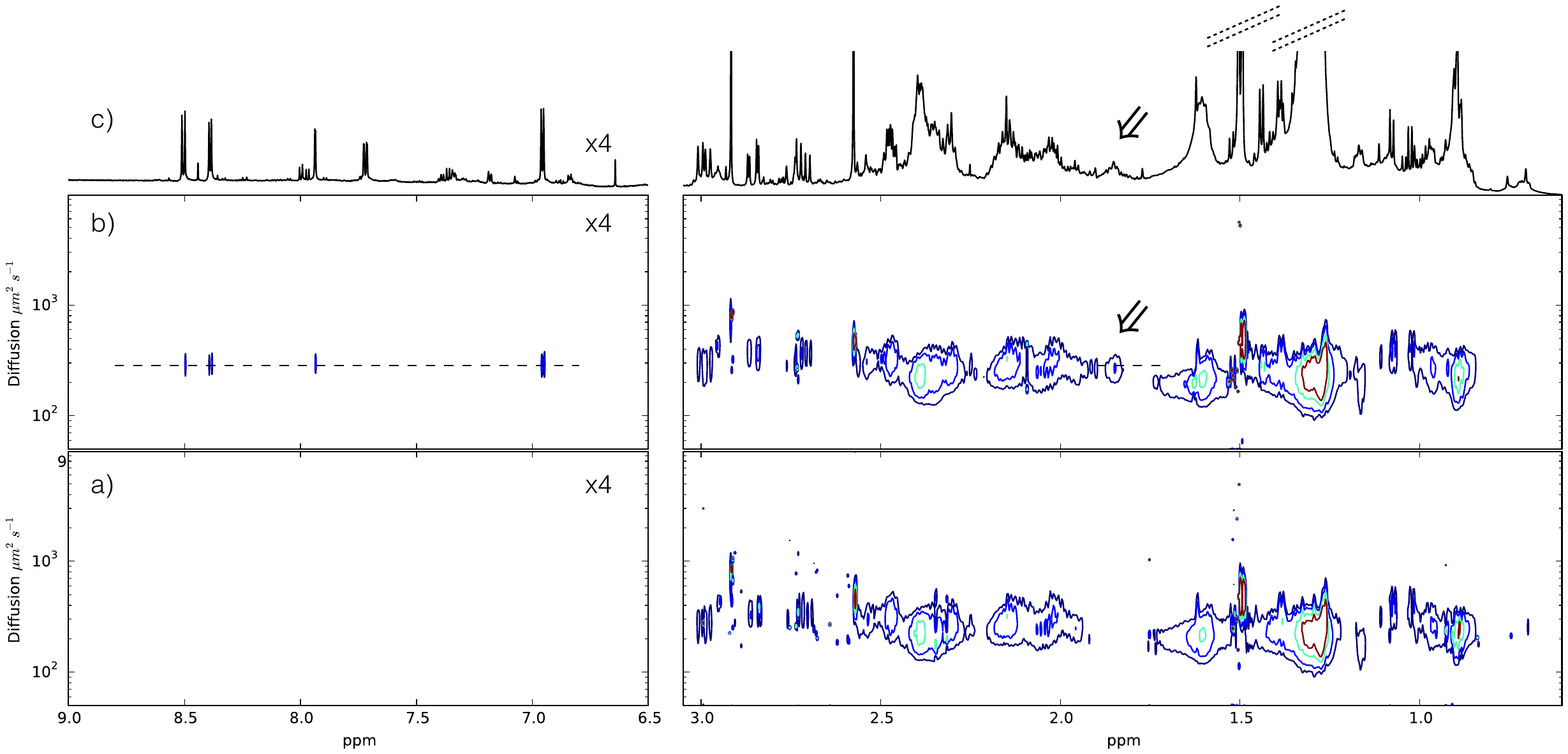}
\caption{Comparison of two DOSY experiments on a brown algae extracts.
a) the aromatic (empty) and aliphatic regions of the same experiment as shown in Fig. \ref{fig:Chlo}, but plotted at a level four times lower;
b) the same brown algae extract with $0.16\,mg\,mL^{-1}$ of chloroquine added, showing the aromatic signals and the methyl signal outlined with an arrow in the spectrum,
the dashed horizontal label is at $285 \mu m^2 s^{-1}$;
c) the 1D spectrum of the brown algae extract with added chloroquine.
In all three spectra, the aromatic panel is plotted four times lower than the aliphatic panel.
}
\label{fig:Chloz}
\end{figure*}

Fig. \ref{fig:Chlo} shows the aliphatic region of the DOSY experiment performed on this algae extract.
Only the more abundant species are visible at this plot level.
This kind of analysis on complex mixture has been extensively used to analyse natural products, plant extract
\cite{Novoa:2011}, or even adulteration of herbal and dietary supplements\cite{Balayssac:2009}.
Here the presence of a particular molecule or family of molecule is characterised by the alignment at the same diffusion coefficient of the characteristic lines located at their corresponding chemical shift positions.
For instance, in Fig. \ref{fig:Chlo}, the polyol and fatty acid signals are outlined.
The fatty acids are certainly partly aggregated in this sample, as indicated by the width of the line, both along the spectral and Laplace axes. 
The DOSY is useful in this context because it provides a high dynamic analysis, where the most intense lines do not "hide" less intense ones.

In Fig. \ref{fig:Chloz}, the chloroquine molecule was added at a low concentration ($1\%\,w/w$) to the same sample.
While the aromatic signals of chloroquine, located in a rather empty region of the spectrum, are easily detected, the aliphatic chain signals fall in crowed region where they are difficult to observe.
The methyl groups that fall at position $1.3$\,ppm and $1.45$\,ppm are completely buried under the fatty acid signals and the diffusion coefficient are not different enough.
In contrast, the signals from the methylene moieties are observed around $1.9$\,ppm, in a relatively free spectral region.
Despite being about $\times100$ smaller than the larger signals (methanol, polyol or fatty acid chains), the signal is well separated in the DOSY spectrum, and is aligned with the aromatic signals.

One can also observed the reproducibility of the PALMA-processed DOSY spectra, as the main features of the spectra are nearly identical for both samples.


\section{Discussion}
The hybrid regularisation proposed in eq \eqref{eq:psi} implements two well known approaches, namely the $\ell_1$ regularisation, which tends to minimise the number of signals required to explain the data, and the MaxEnt regularisation usually presented as a way to maximise the posterior probability of the analysis while preserving the positivity of the retrieved spectra \cite{Chouzenoux_t1t2}.
While both regularisation are well established and based on clear principles, the MaxEnt is known to be somewhat more difficult to implement.
The hybrid regularisation proposed here allows to produce very robust results even in the case of complex signals such as the one presented in the simulated examples.

The constrained problem is solved using a new convex optimisation algorithm, based on the use of proximity operators and a split version of the Douglas-Rachford procedure.
The use of the proximity operators allows to implement a simple incremental step, requiring no inner line-search minimisation step, where the main burden is three applications of the linear operator $\Hb$ or of its generalised inverse $\Bb = (\pmb{I} + \Hb^T \Hb)^{-1}$.
This algorithm allows a rapid convergence even with the hybrid regularisation used here.
In the simulation presented here, an approximate solution is obtained very rapidly (in less than a second).
The results presented in this work were obtained with longer convergences, using typically $10.000$ to $100.000$ iterations, however thanks to the rapidity of the iterative step, this corresponds typically to a few seconds on a laptop.

Because of the constrained approach used here, there is no need to determine a Lagrangian parameter as in most other techniques (sometimes called a \emph{smoothing parameter}).
Nevertheless, the approach requires some parameters, of which the prior $a$ and the noise $\eta$ can readily be estimated from the experimental data-set, using respectively the first point of the decay and an estimate of the noise from a polynomial smoothing of the experimental data-set (see ESI section for details\dag ).
The weight $\lambda$  between the MaxEnt and the $\ell_1$  regularisations embodies a prior assumption on the presence of sparse component in the Laplace spectrum.
In the simulations of a sparse theoretical spectrum (Fig.\ref{fig:Simu}a) $\lambda=0$ corresponding to a pure $\ell_1$ regularisation provides the best reconstruction as expected.
Simulations performed on several wide distributions (Fig.\ref{fig:Simu}b and ESI figures S4, S6, S15, and S16$\dag$) show that the method recovers faithfully the position of the signal and the theoretical profile, for most non-null values of the $\lambda$ parameter, even in the presence of noise.
Results are more contrasted for the sparse spectrum in Fig.\ref{fig:Simu}a, where
the pure MaxEnt analysis ($\lambda=1.0$) presents large features located at the position of the three components, and it can be observed a narrowing of the MaxEnt distribution for decreasing value of $\lambda$.
The results of pure MaxEnt analysis produced by the PALMA algorithm were checked to be equivalent to the results obtained with the original algorithm\cite{Delsuc:1998} based on a fixed point approach, however the convergence is much faster, and processing times are about 10 times shorter for the same results.
It should be recalled that the MaxEnt analysis produces a statistical analysis of the data,
where the final spectrum is the density distribution which maximises the posterior probability of finding a signal\cite{Jaynes:2003,Chouzenoux_t1t2}.
The width of the actual signals can thus be considered as an uncertainty on the position of the monodisperse components, uncertainty that is present in the pure $\ell_1$ case as errors in the position of the lines, but not directly manifest.
These results suffer however from a lack of resolutive power, and the possibility to bias toward a sparser result is certainly a plus.
In a general approach, the optimal value for $\lambda$ should be chosen from assumptions on the data based on explicit previous knowledge, however on a practical point, this is not feasible.
Even on a sample known to be composed solely of monodisperse species, the choice of a null $\lambda$ is problematic.
On the one hand, some polydisperse impurities might be present with the risk of overlooking them as we observe in Fig.\ref{fig:Poly} and Fig.\ref{fig:Chlo};
on the other hand, most instrument imperfections such as temperature drift, gradient non-linearity, phase distortions, convection, etc. will distort the pure exponential decay and create some apparent polydispersity.
Confronted with the same difficulty, \citet{urbanczyk2016monitoring} chose to vary the $p$ parameter of their tailored-ITAMeD algorithm, somewhat similar to $\lambda$ and chose the larger value which allows a minimal residual.
The same approach could easily be used here, however considering the fact that polydisperse samples are correctly analysed for most non-null values of $\lambda$, with the better results obtained for small values,
we suggest using values between $0.01$ and $0.05$ as monodisperse data are well described with these values.

The quality of these results are in sharp contrast with equivalent analyses presented in the literature.
The $\lambda=1$ mode reproduces the classical MaxEnt regularisation\cite{Delsuc:1998} although in less processing time.
The $\lambda=0$ mode, compared to the  $\ell_1$ based ITAMeD approach \cite{urbanczyk2013iterative}, the final resolution resulting from PALMA reconstruction is much higher, as observed in our simulations, in agreement to what has been published.
Intermediate values of $\lambda$, creating a bias of the MaxEnt solution toward more sparse data, produce more resolved spectra, which usually better match the patterns expected in solution NMR.

We recommend using a value of $\lambda$ in the range $0.01$ to $0.05$ for safe results, with the possibility to adapt this value in particular cases (for instance extreme polydispersity or spectral superposition).
We do not recommend to use the pure $\ell_1$ mode ($\lambda=0$), even in the case of monodisperse samples, because the instrumental fluctuations mentioned above certainly disturb this pure behaviour, and because of the difficulty to estimate the uncertainty of the result from a sparse spectrum alone.

\section{Conclusion}
The DOSY experiment holds a special position among the different experiments available to the NMR spectroscopist.
Whilst it provides invaluable information on the size and the interaction of the molecules in solution, it presents an important challenge for its acquisition and its analysis.
he acquired data are usually corrupted by many artefacts produced by the very high sensitivity of this experiment to instrument imperfections such as temperature drift, non-linearity of the gradients and of the detection electronics, phase distortions, convection, etc.
In parallel, while the parameter dependency expressed in the basic evolution equations is extremely simple, it is well established that a simple fit of the data to this equation usually fails in providing a faithful analysis of the data in the general case, and one has to resort to inverse Laplace transform for the analysis step, a problem well known to be of extreme noise sensitivity.
For these reasons, the stability and robustness of the acquisition and processing schemes are of great importance for the quality of DOSY spectroscopy, and many acquisition schemes and many processing procedures have already been proposed in the literature for this purpose.
In this work, we have introduced a general method to solve the inverse problem as found in the analysis of DOSY experiments that we believe provides an unequalled level of quality and robustness in the processing step.

The method is based on a constrained regularisation of the least square problem, and we showed that an hybrid regularisation, combining maximum sparsity and maximum entropy provides optimal results in the general case.
This approach is controlled by a single parameter $\lambda$ weighting between these two criteria, and the results are not very sensitive to the exact value of this parameter as long as extreme values are not chosen.
This method, that we called PALMA, is faster and more robust than previous MaxEnt implementation and provides better results.
It requires only weak assumptions from the user, and can be run in a fully automatic manner.
It has been implemented on a server freely available at \url{http://palma.labo.igbmc.fr}, where users may submit their data-sets for automatic processing.
The code of the algorithm is available at the same address.

\section{Acknowledgements}

This work was supported by the CNRS MASTODONS project under grant 2016TABASCO,
by the Agence Nationale pour la Recherche, grant ANR2014 ONE\_SHOT\_2D\_FT\_ICR.
The authors thank Jean-Christophe Pesquet for the initial idea and for discussion all over this project.
We thank Catherine Vonthron-Sénécheau, Mélanie Bourgeot and Laure Marguerite for the plant extract sample.
We thank Samuel Nicaise for the first version of the WEB server interface
and Julien Seiler for helping in the deployment of the WEB server.
 
\bibliographystyle{rsc}

\providecommand*{\mcitethebibliography}{\thebibliography}
\csname @ifundefined\endcsname{endmcitethebibliography}
{\let\endmcitethebibliography\endthebibliography}{}
\begin{mcitethebibliography}{46}
\providecommand*{\natexlab}[1]{#1}
\providecommand*{\mciteSetBstSublistMode}[1]{}
\providecommand*{\mciteSetBstMaxWidthForm}[2]{}
\providecommand*{\mciteBstWouldAddEndPuncttrue}
  {\def\EndOfBibitem{\unskip.}}
\providecommand*{\mciteBstWouldAddEndPunctfalse}
  {\let\EndOfBibitem\relax}
\providecommand*{\mciteSetBstMidEndSepPunct}[3]{}
\providecommand*{\mciteSetBstSublistLabelBeginEnd}[3]{}
\providecommand*{\EndOfBibitem}{}
\mciteSetBstSublistMode{f}
\mciteSetBstMaxWidthForm{subitem}
{(\emph{\alph{mcitesubitemcount}})}
\mciteSetBstSublistLabelBeginEnd{\mcitemaxwidthsubitemform\space}
{\relax}{\relax}

\bibitem[Stejskal and Tanner(1965)]{stejskal1965spin}
E.~O. Stejskal and J.~E. Tanner, \emph{{J}. {C}hem. {P}hys.}, 1965,
  \textbf{42}, 288--292\relax
\mciteBstWouldAddEndPuncttrue
\mciteSetBstMidEndSepPunct{\mcitedefaultmidpunct}
{\mcitedefaultendpunct}{\mcitedefaultseppunct}\relax
\EndOfBibitem
\bibitem[Sinnaeve(2012)]{Sinnaeve:2012jr}
D.~Sinnaeve, \emph{Concepts in Magnetic Resonance Part A}, 2012, \textbf{40A},
  39--65\relax
\mciteBstWouldAddEndPuncttrue
\mciteSetBstMidEndSepPunct{\mcitedefaultmidpunct}
{\mcitedefaultendpunct}{\mcitedefaultseppunct}\relax
\EndOfBibitem
\bibitem[Johnson(1999)]{johnson1999diffusion}
C.~Johnson, \emph{{P}rog. {N}ucl. {M}agn. {R}eson. {S}pectrosc.}, 1999,
  \textbf{34}, 203--256\relax
\mciteBstWouldAddEndPuncttrue
\mciteSetBstMidEndSepPunct{\mcitedefaultmidpunct}
{\mcitedefaultendpunct}{\mcitedefaultseppunct}\relax
\EndOfBibitem
\bibitem[Morris(2007)]{Morris:2007fy}
G.~A. Morris, \emph{eMagRes}, 2007\relax
\mciteBstWouldAddEndPuncttrue
\mciteSetBstMidEndSepPunct{\mcitedefaultmidpunct}
{\mcitedefaultendpunct}{\mcitedefaultseppunct}\relax
\EndOfBibitem
\bibitem[Price(2009)]{Price:2009}
W.~S. Price, \emph{NMR Studies of Translational Motion: Principles and
  Applications}, Cambridge University Press, Cambride, UK, 2009\relax
\mciteBstWouldAddEndPuncttrue
\mciteSetBstMidEndSepPunct{\mcitedefaultmidpunct}
{\mcitedefaultendpunct}{\mcitedefaultseppunct}\relax
\EndOfBibitem
\bibitem[Callaghan(2011)]{Callaghan:2011}
P.~T. Callaghan, \emph{Translational dynamics and magnetic resonance:
  principles of pulsed gradient spin echo NMR}, Oxford University Press,
  Oxford, UK, 2011\relax
\mciteBstWouldAddEndPuncttrue
\mciteSetBstMidEndSepPunct{\mcitedefaultmidpunct}
{\mcitedefaultendpunct}{\mcitedefaultseppunct}\relax
\EndOfBibitem
\bibitem[Stilbs \emph{et~al.}(1996)Stilbs, Paulsen, and Griffiths]{Stilbs:1996}
P.~Stilbs, K.~Paulsen and P.~Griffiths, \emph{J. Phys. Chem.}, 1996,
  \textbf{100}, 8180--8189\relax
\mciteBstWouldAddEndPuncttrue
\mciteSetBstMidEndSepPunct{\mcitedefaultmidpunct}
{\mcitedefaultendpunct}{\mcitedefaultseppunct}\relax
\EndOfBibitem
\bibitem[{Van Gorkom} and Hancewicz(1998)]{Gorkom:1998}
L.~{Van Gorkom} and T.~Hancewicz, \emph{J. Magn. Reson.}, 1998, \textbf{130},
  125--130\relax
\mciteBstWouldAddEndPuncttrue
\mciteSetBstMidEndSepPunct{\mcitedefaultmidpunct}
{\mcitedefaultendpunct}{\mcitedefaultseppunct}\relax
\EndOfBibitem
\bibitem[Windig and Antalek(1997)]{Windig:1997}
W.~Windig and B.~Antalek, \emph{Chemom. Intell. Lab. Syst.}, 1997, \textbf{37},
  241--254\relax
\mciteBstWouldAddEndPuncttrue
\mciteSetBstMidEndSepPunct{\mcitedefaultmidpunct}
{\mcitedefaultendpunct}{\mcitedefaultseppunct}\relax
\EndOfBibitem
\bibitem[Armstrong \emph{et~al.}(2003)Armstrong, Loening, Curtis, Shaka, and
  Mandelshtam]{Armstrong:2003}
G.~Armstrong, N.~Loening, J.~Curtis, A.~Shaka and V.~Mandelshtam, \emph{J.
  Magn. Reson.}, 2003, \textbf{163}, 139--148\relax
\mciteBstWouldAddEndPuncttrue
\mciteSetBstMidEndSepPunct{\mcitedefaultmidpunct}
{\mcitedefaultendpunct}{\mcitedefaultseppunct}\relax
\EndOfBibitem
\bibitem[Huo \emph{et~al.}(2004)Huo, Wehrens, and Buydens]{Huo:2004uj}
R.~Huo, R.~Wehrens and L.~Buydens, \emph{J Magn Reson}, 2004, \textbf{169},
  257--269\relax
\mciteBstWouldAddEndPuncttrue
\mciteSetBstMidEndSepPunct{\mcitedefaultmidpunct}
{\mcitedefaultendpunct}{\mcitedefaultseppunct}\relax
\EndOfBibitem
\bibitem[Nilsson and Morris(2008)]{Nilsson:2008}
M.~Nilsson and G.~Morris, \emph{Anal. Chem.}, 2008, \textbf{80},
  3777--3782\relax
\mciteBstWouldAddEndPuncttrue
\mciteSetBstMidEndSepPunct{\mcitedefaultmidpunct}
{\mcitedefaultendpunct}{\mcitedefaultseppunct}\relax
\EndOfBibitem
\bibitem[Stilbs(2010)]{Stilbs:2010}
P.~Stilbs, \emph{J. Magn. Reson.}, 2010, \textbf{207}, 332--336\relax
\mciteBstWouldAddEndPuncttrue
\mciteSetBstMidEndSepPunct{\mcitedefaultmidpunct}
{\mcitedefaultendpunct}{\mcitedefaultseppunct}\relax
\EndOfBibitem
\bibitem[Martini \emph{et~al.}(2013)Martini, Mandelshtam, Morris, Colbourne,
  and Nilsson]{Martini:2013ge}
B.~R. Martini, V.~A. Mandelshtam, G.~A. Morris, A.~A. Colbourne and M.~Nilsson,
  \emph{J Magn Reson}, 2013, \textbf{234}, 125--134\relax
\mciteBstWouldAddEndPuncttrue
\mciteSetBstMidEndSepPunct{\mcitedefaultmidpunct}
{\mcitedefaultendpunct}{\mcitedefaultseppunct}\relax
\EndOfBibitem
\bibitem[D.Nuzillard \emph{et~al.}(1998)D.Nuzillard, S.Bourgand, and
  J.-M.Nuzillard]{Nuzillard:1998}
D.Nuzillard, S.Bourgand and J.-M.Nuzillard, \emph{J.Magn.Reson.}, 1998,
  \textbf{133}, 358--363\relax
\mciteBstWouldAddEndPuncttrue
\mciteSetBstMidEndSepPunct{\mcitedefaultmidpunct}
{\mcitedefaultendpunct}{\mcitedefaultseppunct}\relax
\EndOfBibitem
\bibitem[Naanaa and Nuzillard(2005)]{Nuzillard:2005}
W.~Naanaa and J.-M. Nuzillard, \emph{Sign. Process.}, 2005, \textbf{85},
  1711--1722\relax
\mciteBstWouldAddEndPuncttrue
\mciteSetBstMidEndSepPunct{\mcitedefaultmidpunct}
{\mcitedefaultendpunct}{\mcitedefaultseppunct}\relax
\EndOfBibitem
\bibitem[Colbourne \emph{et~al.}(2011)Colbourne, Morris, and
  Nilsson]{Colbourne:2011}
A.~Colbourne, G.~Morris and M.~Nilsson, \emph{J. Am. Chem. Soc.}, 2011,
  \textbf{133}, 7640--7643\relax
\mciteBstWouldAddEndPuncttrue
\mciteSetBstMidEndSepPunct{\mcitedefaultmidpunct}
{\mcitedefaultendpunct}{\mcitedefaultseppunct}\relax
\EndOfBibitem
\bibitem[Toumi \emph{et~al.}(2013)Toumi, Torr{\'e}sani, and
  Caldarelli]{Toumi:2013iy}
I.~Toumi, B.~Torr{\'e}sani and S.~Caldarelli, \emph{Anal Chem}, 2013,
  \textbf{85}, 11344--11351\relax
\mciteBstWouldAddEndPuncttrue
\mciteSetBstMidEndSepPunct{\mcitedefaultmidpunct}
{\mcitedefaultendpunct}{\mcitedefaultseppunct}\relax
\EndOfBibitem
\bibitem[Vi{\'e}ville \emph{et~al.}(2011)Vi{\'e}ville, Tanty, and
  Delsuc]{vieville2011polydispersity}
J.~Vi{\'e}ville, M.~Tanty and M.-A. Delsuc, \emph{{J}. {M}agn. {R}eson}, 2011,
  \textbf{212}, 169--173\relax
\mciteBstWouldAddEndPuncttrue
\mciteSetBstMidEndSepPunct{\mcitedefaultmidpunct}
{\mcitedefaultendpunct}{\mcitedefaultseppunct}\relax
\EndOfBibitem
\bibitem[Williamson \emph{et~al.}(2016)Williamson, Nyd{\'e}n, and
  R{\"o}ding]{Williamson:2016ch}
N.~H. Williamson, M.~Nyd{\'e}n and M.~R{\"o}ding, \emph{J Magn Reson}, 2016,
  \textbf{267}, 54--62\relax
\mciteBstWouldAddEndPuncttrue
\mciteSetBstMidEndSepPunct{\mcitedefaultmidpunct}
{\mcitedefaultendpunct}{\mcitedefaultseppunct}\relax
\EndOfBibitem
\bibitem[Provencher(1982)]{provencher1982constrained}
S.~W. Provencher, \emph{{C}omput. {P}hys.}, 1982, \textbf{27}, 213--227\relax
\mciteBstWouldAddEndPuncttrue
\mciteSetBstMidEndSepPunct{\mcitedefaultmidpunct}
{\mcitedefaultendpunct}{\mcitedefaultseppunct}\relax
\EndOfBibitem
\bibitem[Nityananda and Narayan(1982)]{Nityananda82}
R.~Nityananda and R.~Narayan, \emph{{A}stron. {A}strophys.}, 1982, \textbf{3},
  year\relax
\mciteBstWouldAddEndPuncttrue
\mciteSetBstMidEndSepPunct{\mcitedefaultmidpunct}
{\mcitedefaultendpunct}{\mcitedefaultseppunct}\relax
\EndOfBibitem
\bibitem[Delsuc and Malliavin(1998)]{Delsuc:1998}
M.-A. Delsuc and T.~E. Malliavin, \emph{Anal. Chem.}, 1998, \textbf{70},
  2146--2148\relax
\mciteBstWouldAddEndPuncttrue
\mciteSetBstMidEndSepPunct{\mcitedefaultmidpunct}
{\mcitedefaultendpunct}{\mcitedefaultseppunct}\relax
\EndOfBibitem
\bibitem[Skilling and Bryan(1984)]{Skilling:1984ti}
J.~Skilling and R.~Bryan, \emph{Mon. Not. R. Astron. Soc.}, 1984, \textbf{211},
  111\relax
\mciteBstWouldAddEndPuncttrue
\mciteSetBstMidEndSepPunct{\mcitedefaultmidpunct}
{\mcitedefaultendpunct}{\mcitedefaultseppunct}\relax
\EndOfBibitem
\bibitem[Urba{\'n}czyk \emph{et~al.}(2013)Urba{\'n}czyk, Bernin, Kozminski, and
  Kazimierczuk]{urbanczyk2013iterative}
M.~Urba{\'n}czyk, D.~Bernin, W.~Kozminski and K.~Kazimierczuk, \emph{Anal.
  Chem.}, 2013, \textbf{85}, 1828--1833\relax
\mciteBstWouldAddEndPuncttrue
\mciteSetBstMidEndSepPunct{\mcitedefaultmidpunct}
{\mcitedefaultendpunct}{\mcitedefaultseppunct}\relax
\EndOfBibitem
\bibitem[Urba{\'n}czyk \emph{et~al.}(2016)Urba{\'n}czyk, Bernin, Czuro{\'n},
  and Kazimierczuk]{urbanczyk2016monitoring}
M.~Urba{\'n}czyk, D.~Bernin, A.~Czuro{\'n} and K.~Kazimierczuk, \emph{Analyst},
  2016, \textbf{141}, 1745--1752\relax
\mciteBstWouldAddEndPuncttrue
\mciteSetBstMidEndSepPunct{\mcitedefaultmidpunct}
{\mcitedefaultendpunct}{\mcitedefaultseppunct}\relax
\EndOfBibitem
\bibitem[{Kazimierczuk, Krzysztof} and {Orekhov, Vladislav
  Yu}(2011)]{Kazimierczuk:2011jy}
{Kazimierczuk, Krzysztof} and {Orekhov, Vladislav Yu}, \emph{Angew. Chem. Int.
  Ed.}, 2011, \textbf{50}, 5556--5559\relax
\mciteBstWouldAddEndPuncttrue
\mciteSetBstMidEndSepPunct{\mcitedefaultmidpunct}
{\mcitedefaultendpunct}{\mcitedefaultseppunct}\relax
\EndOfBibitem
\bibitem[Xu and Zhang(2013)]{xu2013trust}
K.~Xu and S.~Zhang, \emph{{A}nal. {C}hem.}, 2013, \textbf{86}, 592--599\relax
\mciteBstWouldAddEndPuncttrue
\mciteSetBstMidEndSepPunct{\mcitedefaultmidpunct}
{\mcitedefaultendpunct}{\mcitedefaultseppunct}\relax
\EndOfBibitem
\bibitem[Idier(2008)]{Idier:2008}
\emph{Bayesian Approach to Inverse Problems}, ed. J.~Idier, ISTE, 2008\relax
\mciteBstWouldAddEndPuncttrue
\mciteSetBstMidEndSepPunct{\mcitedefaultmidpunct}
{\mcitedefaultendpunct}{\mcitedefaultseppunct}\relax
\EndOfBibitem
\bibitem[Elad(2010)]{eladCS}
M.~Elad, \emph{Sparse and Redundant Representations}, Springer, New York, NY,
  2010\relax
\mciteBstWouldAddEndPuncttrue
\mciteSetBstMidEndSepPunct{\mcitedefaultmidpunct}
{\mcitedefaultendpunct}{\mcitedefaultseppunct}\relax
\EndOfBibitem
\bibitem[Bauschke and Combettes(2011)]{bauschke2011convex}
H.~H. Bauschke and P.~L. Combettes, \emph{Convex analysis and monotone operator
  theory in Hilbert spaces}, Springer, New York, NY, 2011\relax
\mciteBstWouldAddEndPuncttrue
\mciteSetBstMidEndSepPunct{\mcitedefaultmidpunct}
{\mcitedefaultendpunct}{\mcitedefaultseppunct}\relax
\EndOfBibitem
\bibitem[Beck and Teboulle(2009)]{Beck09}
A.~Beck and M.~Teboulle, \emph{{S}IAM. {J}. {I}maging. {S}ci.}, 2009,
  \textbf{2}, 183--202\relax
\mciteBstWouldAddEndPuncttrue
\mciteSetBstMidEndSepPunct{\mcitedefaultmidpunct}
{\mcitedefaultendpunct}{\mcitedefaultseppunct}\relax
\EndOfBibitem
\bibitem[Cherni \emph{et~al.}(2016)Cherni, Chouzenoux, and
  Delsuc]{CherniISIVC2016}
A.~Cherni, E.~Chouzenoux and M.-A. Delsuc, Proc. 8th Int. Symp. Signal, Image,
  Video and Commun., 2016, pp. {x--x+6}\relax
\mciteBstWouldAddEndPuncttrue
\mciteSetBstMidEndSepPunct{\mcitedefaultmidpunct}
{\mcitedefaultendpunct}{\mcitedefaultseppunct}\relax
\EndOfBibitem
\bibitem[Corless \emph{et~al.}(1996)Corless, Gonnet, Hare, Jeffrey, and
  Knuth]{corless1996lambertw}
R.~M. Corless, G.~H. Gonnet, D.~E.~G. Hare, D.~J. Jeffrey and D.~E. Knuth,
  \emph{{A}dv. {C}omput. {M}ath}, 1996, \textbf{5}, 329--359\relax
\mciteBstWouldAddEndPuncttrue
\mciteSetBstMidEndSepPunct{\mcitedefaultmidpunct}
{\mcitedefaultendpunct}{\mcitedefaultseppunct}\relax
\EndOfBibitem
\bibitem[Combettes and Pesquet(2011)]{combettes2011proximal}
P.~L. Combettes and J.-C. Pesquet, in \emph{Fixed-Point Algorithms for Inverse
  Problems in Science and Engineering}, ed. H.~H. Bauschke, S.~R. Burachik,
  L.~P. Combettes, V.~Elser, R.~D. Luke and H.~Wolkowicz, Springer, New York,
  NY, 2011, pp. 185--212\relax
\mciteBstWouldAddEndPuncttrue
\mciteSetBstMidEndSepPunct{\mcitedefaultmidpunct}
{\mcitedefaultendpunct}{\mcitedefaultseppunct}\relax
\EndOfBibitem
\bibitem[Pustelnik \emph{et~al.}(2011)Pustelnik, Chaux, and
  Pesquet]{Pustelnik_2011_hybrid_regularization_restoration}
N.~Pustelnik, C.~Chaux and J.-C. Pesquet, \emph{{IEEE} {T}rans. {I}mage
  {P}rocess.}, 2011, \textbf{20}, 2450--2462\relax
\mciteBstWouldAddEndPuncttrue
\mciteSetBstMidEndSepPunct{\mcitedefaultmidpunct}
{\mcitedefaultendpunct}{\mcitedefaultseppunct}\relax
\EndOfBibitem
\bibitem[Combettes and Pesquet(December 2008)]{Combettes2008PPXA}
P.~L. Combettes and J.-C. Pesquet, \emph{Inverse problems}, December 2008,
  \textbf{24}, 564--574\relax
\mciteBstWouldAddEndPuncttrue
\mciteSetBstMidEndSepPunct{\mcitedefaultmidpunct}
{\mcitedefaultendpunct}{\mcitedefaultseppunct}\relax
\EndOfBibitem
\bibitem[Combettes and Pesquet(December 2007)]{Combettes2007DouglasRachford}
P.~L. Combettes and J.-C. Pesquet, \emph{IEEE J. Sel. Topics Signal Process},
  December 2007, \textbf{1}, 564--574\relax
\mciteBstWouldAddEndPuncttrue
\mciteSetBstMidEndSepPunct{\mcitedefaultmidpunct}
{\mcitedefaultendpunct}{\mcitedefaultseppunct}\relax
\EndOfBibitem
\bibitem[Vonthron-Sénécheau \emph{et~al.}(2011)Vonthron-Sénécheau, Kaiser,
  Devambez, Vastel, Mussio, and Rusig]{Vonthron:2011}
C.~Vonthron-Sénécheau, M.~Kaiser, I.~Devambez, A.~Vastel, I.~Mussio and A.-M.
  Rusig, \emph{Mar. Drugs}, 2011, \textbf{9}, 922--933\relax
\mciteBstWouldAddEndPuncttrue
\mciteSetBstMidEndSepPunct{\mcitedefaultmidpunct}
{\mcitedefaultendpunct}{\mcitedefaultseppunct}\relax
\EndOfBibitem
\bibitem[Jerschow and M{\"u}ller(1997)]{Jerschow:1997vm}
A.~Jerschow and N.~M{\"u}ller, \emph{J Magn Reson}, 1997, \textbf{125},
  372--375\relax
\mciteBstWouldAddEndPuncttrue
\mciteSetBstMidEndSepPunct{\mcitedefaultmidpunct}
{\mcitedefaultendpunct}{\mcitedefaultseppunct}\relax
\EndOfBibitem
\bibitem[Assemat \emph{et~al.}(2010)Assemat, Coutouly, Hajjar, and
  Delsuc]{assemat:2010ti}
O.~Assemat, M.-A. Coutouly, R.~Hajjar and M.-A. Delsuc, \emph{C. R. Chim},
  2010, \textbf{13}, 412--415\relax
\mciteBstWouldAddEndPuncttrue
\mciteSetBstMidEndSepPunct{\mcitedefaultmidpunct}
{\mcitedefaultendpunct}{\mcitedefaultseppunct}\relax
\EndOfBibitem
\bibitem[Chiron \emph{et~al.}(2016)Chiron, Coutouly, Starck, Rolando, and
  Delsuc]{spike}
L.~Chiron, M.-A. Coutouly, J.-P. Starck, C.~Rolando and M.-A. Delsuc,
  \emph{arXiv}, 2016, \textbf{1608.06777}, 1--13\relax
\mciteBstWouldAddEndPuncttrue
\mciteSetBstMidEndSepPunct{\mcitedefaultmidpunct}
{\mcitedefaultendpunct}{\mcitedefaultseppunct}\relax
\EndOfBibitem
\bibitem[Novoa-Carballal \emph{et~al.}(2011)Novoa-Carballal, Fernandez-Megia,
  Jimenez, and Riguera]{Novoa:2011}
R.~Novoa-Carballal, E.~Fernandez-Megia, C.~Jimenez and R.~Riguera, \emph{Nat.
  Prod. Rep.}, 2011, \textbf{28}, 78--98\relax
\mciteBstWouldAddEndPuncttrue
\mciteSetBstMidEndSepPunct{\mcitedefaultmidpunct}
{\mcitedefaultendpunct}{\mcitedefaultseppunct}\relax
\EndOfBibitem
\bibitem[Balayssac \emph{et~al.}(2009)Balayssac, Trefi, Gilard, Malet-Martino,
  Martino, and Delsuc]{Balayssac:2009}
S.~Balayssac, S.~Trefi, V.~Gilard, M.~Malet-Martino, R.~Martino and M.-A.
  Delsuc, \emph{J Pharm Biomed Anal}, 2009, \textbf{50}, 602--612\relax
\mciteBstWouldAddEndPuncttrue
\mciteSetBstMidEndSepPunct{\mcitedefaultmidpunct}
{\mcitedefaultendpunct}{\mcitedefaultseppunct}\relax
\EndOfBibitem
\bibitem[Chouzenoux \emph{et~al.}(2010)Chouzenoux, Moussaoui, Idier, and
  Mariette]{Chouzenoux_t1t2}
E.~Chouzenoux, S.~Moussaoui, J.~Idier and F.~Mariette, \emph{IEEE Transactions
  on Signal Processing}, 2010, \textbf{58}, 6040--6051\relax
\mciteBstWouldAddEndPuncttrue
\mciteSetBstMidEndSepPunct{\mcitedefaultmidpunct}
{\mcitedefaultendpunct}{\mcitedefaultseppunct}\relax
\EndOfBibitem
\bibitem[Jaynes(2003)]{Jaynes:2003}
E.~Jaynes, \emph{Probability Theory: The Logic of Science}, Cambridge
  University Press., Cambridge, UK, 2003\relax
\mciteBstWouldAddEndPuncttrue
\mciteSetBstMidEndSepPunct{\mcitedefaultmidpunct}
{\mcitedefaultendpunct}{\mcitedefaultseppunct}\relax
\EndOfBibitem
\end{mcitethebibliography}
\providecommand*{\mcitethebibliography}{\thebibliography}
\csname @ifundefined\endcsname{endmcitethebibliography}
{\let\endmcitethebibliography\endthebibliography}{}



\section*{Supplementary material}

# Support Information for: PALMA, an improved algorithm for the DOSY signal processing

Afef Cherni[a,b], Emilie Chouzenoux[b], Marc-André Delsuc[*,a]

# Contents

[a]: Institut de Génétique et de Biologie Moléculaire et Cellulaire (IGBMC), INSERM U596, CNRS UMR 7104 Université de Strasbourg, 67404 Illkirch-Graffenstaden, France.

[b]: Université Paris-Est, LIGM (UMR 8049), CNRS, ENPC, ESIEE Paris, UPEM, Marne-la-Vallée, France.

[*] madelsuc@unistra.fr

# 1 Variational formulation

A well-known and very efficient strategy for solving ill-posed inverse problems is to adopt a penalization approach that provides an estimate $\hat{X} \in \mathbb{R}^N$ of the original signal $X \in R^N$, that is the solution of the constrained minimization problem:

$$\underset{X \in \mathbb{R}^N}{\text{minimize}} \ \Psi(X) \quad \text{subject to} \quad \|\boldsymbol{H}X - Y\| \leq \eta, \tag{1}$$

where $\Psi$ is the regularization function and $\eta$ is an estimate of the experimental noise. In this work, we choose to define $\Psi$ for every $X \in \mathbb{R}^N$ and $\lambda \in [0,1]$, as:

$$\Psi(X) = \lambda \operatorname{ent}(X, a) + (1 - \lambda)\ell_1(X), \tag{2}$$

where the Shannon entropy with a flat prior $a > 0$ and the $\ell_1$ norm are defined, respectively, as:

$$(\forall X = (x_n)_{1 \leq n \leq N} \in \mathbb{R}^N) \quad \operatorname{ent}(X, a) = \begin{cases} \sum_{n=1}^N \frac{x_n}{a} \log\left(\frac{x_n}{a}\right) & \text{if } x_n > 0 \\ 0 & \text{if } x_n = 0 \\ +\infty & \text{elsewhere} \end{cases}$$

and

$$(\forall X = (x_n)_{1 \leq n \leq N} \in \mathbb{R}^N) \quad \ell_1(X) = \sum_{n=1}^N |x_n|.$$

# 2 Proximity operator

To solve Problem (1), we propose to rely on a proximal optimization method, that makes use of the so-called proximity operator. Since $\Psi$ is convex on $\mathbb{R}^N$, its proximity operator at a given point $X \in \mathbb{R}^N$ is defined as the unique minimizer of $\frac{1}{2}\|\cdot - X\|^2 + \Psi$. Since $\Psi$ in (2) takes a separable form, i.e. it can be written as:

$$(\forall X = (x_n)_{1 \leq n \leq N} \in \mathbb{R}^N) \quad \Psi(X) = \sum_{n=1}^N \psi(x_n), \tag{3}$$

its proximity operator is given by:

$$\operatorname{prox}_\Psi(x) = (p(x_n))_{1 \leq n \leq N}. \tag{4}$$

Hereabove, for every $n \in \{1, \ldots, N\}$, $p(x_n) \in \mathbb{R}$ is the unique minimizer of:

$$\phi : u \mapsto \frac{1}{2}(u - x_n)^2 + \psi(u). \tag{5}$$

Let us first consider the case when $\lambda \in ]0, 1]$. Then,

$$\dot{\phi}(u) = 0 \Leftrightarrow u - x_n + \frac{\lambda}{a} \log(u) + \frac{\lambda}{a} - \frac{\lambda}{a} \log(a) + 1 - \lambda = 0$$

$$\Leftrightarrow \log(u) = \frac{a}{\lambda} \left( -u + x - \frac{\lambda}{a} + \frac{\lambda}{a} \log(a) - 1 + \lambda \right)$$

$$\Leftrightarrow u = \exp\left( -\frac{a}{\lambda} u + \frac{ax - a(1-\lambda)}{\lambda} + \log(a) - 1 \right).$$

Finally:

$$u = \frac{\lambda}{a} \mathcal{W}\left( \frac{a}{\lambda} \exp\left( \frac{ax - a(1-\lambda)}{\lambda} + \log(a) - 1 \right) \right). \tag{6}$$

In the above expression, $\mathcal{W}$ states for the Lambert function, also called Omega, defined as the inverse function of $f : x \to x \exp(x)$ for all $x \in \mathbb{C}$:

$$z = x \exp(x) \Leftrightarrow x = \mathcal{W}(z).$$

If $\lambda = 0$, we have:

$$x_n - u \in \partial |u| \Longleftrightarrow x_n - u \in \begin{cases} \operatorname{sign}(u) & \text{if } u \neq 0 \\ [-1, 1] & \text{elsewhere} \end{cases}$$

$$\Longleftrightarrow u = \begin{cases} x_n - 1 & \text{if } u > 1 \\ 0 & \text{if } u \in [-1, 1] \\ x_n + 1 & \text{elsewhere.} \end{cases}$$

Finally, we can conclude that $(\forall X = (x_n)_{1 \leq n \leq N} \in R^N)$, $\operatorname{prox}_\Psi(X)$ is given by (4), where, for all $n \in \{1, \ldots, N\}$,

$$p(x_n) = \begin{cases} \frac{\lambda}{a} \mathcal{W}\left( \frac{a}{\lambda} \exp\left( \frac{ax_n - a(1-\lambda)}{\lambda} + \log(a) - 1 \right) \right) & \text{if } \lambda \in ]0, 1] \\ \operatorname{sign}(x_n) \left( \max(|x_n| - 1, 0) \right) & \text{if } \lambda = 0. \end{cases} \tag{7}$$

## 3 Asymptotic development

Let us emphasize that the Lambert function has many interesting properties. In particular:

$$\mathcal{W}(e^x) \xrightarrow[x \to +\infty]{} x \left( 1 - \frac{\log(x)}{1 + x} \right) \tag{8}$$

This asymptotic development is of main interest in our context. Indeed, when $\lambda \in ]0, 1]$, for all $n \in \{1, \ldots, N\}$,

$$p(x_n) = \frac{\lambda}{a} \mathcal{W}(\exp(c_n)), \tag{9}$$

with $c_n = \frac{ax_n - a(1-\lambda)}{\lambda} - 1 + 2\log(a) - \log(\lambda)$. When $c_n$ is large (typically $c_n > 10^2$), standard numerical implementation of the Lambert function yields infinity as an output.

We thus propose to use, for large $c_n$, the following approximation of the proximity operator, which is a consequence of the asymptotic development (8):

$$p(x_n) = \frac{\lambda}{a}\left(c_n - \frac{c_n}{1+c_n}\log(c_n)\right). \tag{10}$$

Figure 8 illustrates relative error of the approximation, when $\lambda = a = 1$.

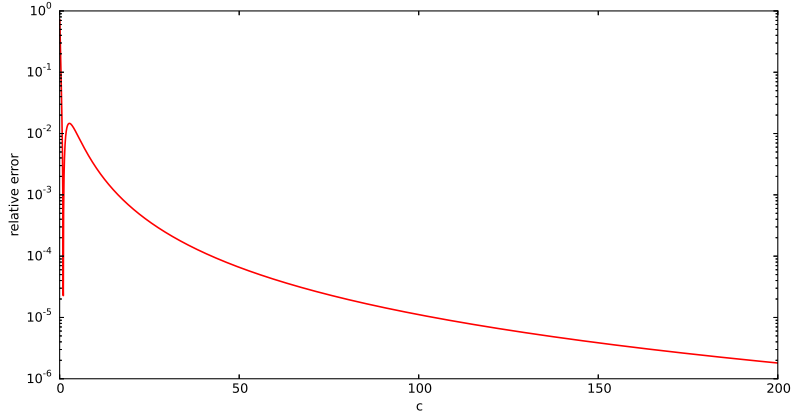

Figure S1: Relative error value between $\mathcal{W}(e^{c_n})$ and the approximation of $p(x_n)$ in 9. The relative error becomes lower than $10^{-3}$ for $c_n > 20$.

## 4 PALMA algorithm

Using the PPXA+ algorithm, we propose a new proximal algorithm to solve (1) that makes use of the proximal formula (7). The so-called PALMA algorithm, standing for "**P**roximal **A**lgorithm for **L**$_1$ combined with **MA**xent prior", is given below:

**Initialization**
$V^{(0,1)} \in \mathbb{R}^N, V^{(0,2)} \in \mathbb{R}^M$
$X^{(0)} = (\mathrm{I}_N + \boldsymbol{H}^\top \boldsymbol{H})^{-1}(V^{(0,1)} + \boldsymbol{H}^\top V^{(0,2)})$
$\boldsymbol{B} = (\mathrm{I}_N + \boldsymbol{H}^\top \boldsymbol{H})^{-1}$
$\gamma \in (0, 2)$,
**Minimization**
For $k = 0, 1, \ldots$
$\quad Z^{(k,1)} = \mathrm{prox}_\Psi(V^{(k,1)})$
$\quad Z^{(k,2)} = \mathrm{proj}_{\|\cdot - Y\| \leq \eta}(V^{(k,2)})$
$\quad U^{(k)} = \boldsymbol{B}(Z^{(k,1)} + \boldsymbol{H}^\top Z^{(k,2)})$
$\quad X^{(k+1)} = X^{(k)} + \gamma(U^{(k)} - X^{(k)})$
$\quad V^{(k+1,1)} = V^{(k,1)} + \gamma\left(2U^{(k)} - X^{(k)} - Z^{(k,1)}\right)$
$\quad V^{(k+1,2)} = V^{(k,2)} + \gamma\left(\boldsymbol{H}(2U^{(k)} - X^{(k)}) - Z^{(k,2)}\right)$

where $I_N$ is the identity matrix of $\mathbb{R}^N$ and the projection $\text{proj}_{\|\cdot-Y\|\leq\eta}$ is defined as follows:

$$\text{proj}_{\|\cdot-Y\|\leq\eta}(Z) = Z + (Z-Y)\min\left(\frac{\eta}{\|Z-Y\|},1\right) - Y \quad (\forall(Y,Z)\in(\mathbb{R}^N)^2).$$

# 5 Choice of processing parameters

As it is detailed in the previous theoretical section, the PALMA algorithm is controlled by several scaling parameters which have to be adapted to the current problem.

## 5.1 Choice of $a$

The entropy approach requires an expression of the prior knowledge on the system, which in the present case corresponds to *a priori* spectrum in the absence of experimental evidences. A general expression is thus a flat spectrum of intensity $a$, the value of which has to be adapted to the scaling of the experimental measurement. A natural approach consists in estimating the area under the signal expected signal: $\sum_n x_n$ and scaling the prior with this value. Due to the properties of the Laplace transform, we have chosen here $a = y_0 \approx \sum_n x_n$.

## 5.2 Choice of $\eta$

If we assume that the experimental values are tainted by an additive random noise: $y_m = \hat{y}_m + \varepsilon_m$, and if this noise is supposed to be centered and Gaussian i.i.d. with variance $\sigma^2$, then we can expect the residual of the fit to be $\eta \approx \sigma\sqrt{M}$. The value of $\sigma$ can be measured here either from additional measures or from a simple polynomial fit of the $y_m$ curve as it is done in the current implementation. The case of a correlated Gaussian noise $\varepsilon$, with covariance matrix $\Sigma$, is encompassed by our method. Indeed, the least squares constraint becomes:

$$(\forall X \in \mathbb{R}^N) \quad \left((\boldsymbol{H}X - Y)^\top \Sigma^{-1}(\boldsymbol{H}X - Y)\right)^{1/2} \leq \eta,$$

which is equivalent to $\|\boldsymbol{H}X - Y\|_2 \leq \eta$, up to any change of variable with the form $\boldsymbol{H} \longleftarrow \theta\Sigma^{-1/2}\boldsymbol{H}$ and $Y \longleftarrow \theta\Sigma^{-1/2}Y$, $\eta \longleftarrow \theta\eta$, for some $\theta > 0$. In order to facilitate the choice of $\eta$, we propose to take $\theta = \sigma_{\max}$, i.e. the maximal singular value of $\Sigma$, so that a suitable choice for $\eta$ is $\sigma_{\max}\sqrt{M}$.

## 5.3 Choice of $\lambda$

In order to evaluate the influence of the $\lambda$ factor on the balance between sparsity and the amount of information in the processed signal, we launch our PALMA algorithm for different signal of monodisperse and polysdisperse species (A, B, C1 and C2) and we trace on Figures S2 to S6, the recovered signal by varying $\lambda$ from 0 to 1.

The associated reconstruction quality, measured in terms of signal to noise ratio (SNR) $= 10 \log 10 \left( \frac{\|X\|}{\|\hat{X}-X\|} \right)$ of each tested case, is reported in Table S1.
**Larger value of quality of reconstruction correspondents to the best reconstruction.**

• **Signal A**

Signal A consists in three monodisperse components with diffusion coefficients $16\,\mu m^2/s$, $63\,\mu m^2/s$, and $230\,\mu m^2/s$, with respective intensities $1.0, 0.33$ and $0.66$. This data-set is equivalent to the simulation used in *Kazimierczuk et al.*

• **Signal B**

Signal B is a wide monodisperse distribution, simulated as a symmetric log-normal distribution centered at $35\,\mu m^2/s$ and with variance 25.

• **Signals C1 & C2**

C1 and C2 signals are asymmetric distributions built from 15 log-normal components, ranging from 18 to $85\,\mu m^2/s$, with intensities ranging from 0.1 to $10\,s$. They have PDI estimated respectively to 1.79 and 1.32.

For C1, the intensities are chosen as $10.0/(1.4^i), i = 1, \ldots, 15$, and for C2, as $10.0/(1.4^{15-i}), i = 1, \ldots, 15$.

In all simulations, an additive zero-mean white Gaussian noise and standard deviation $\sigma$ equals $0.1\%$ of the initial point of $\boldsymbol{H}X$ was added to the observed data. The number of measurement is $M = 64$ and the original signal has the dimension $N = 256$.

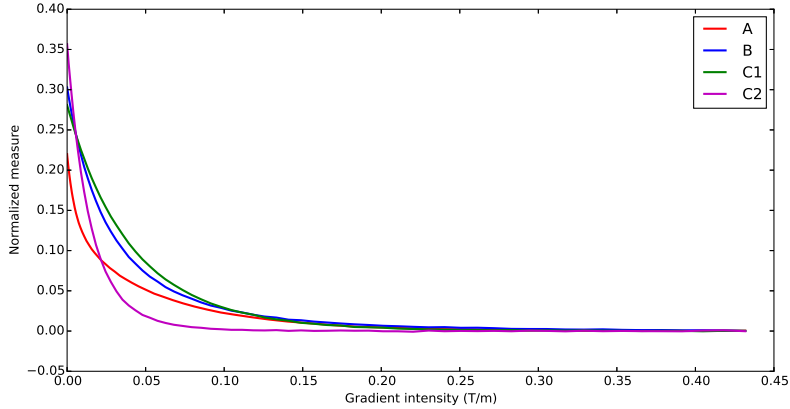

Figure S2: Measurement of A, B, C1 and C2 signals

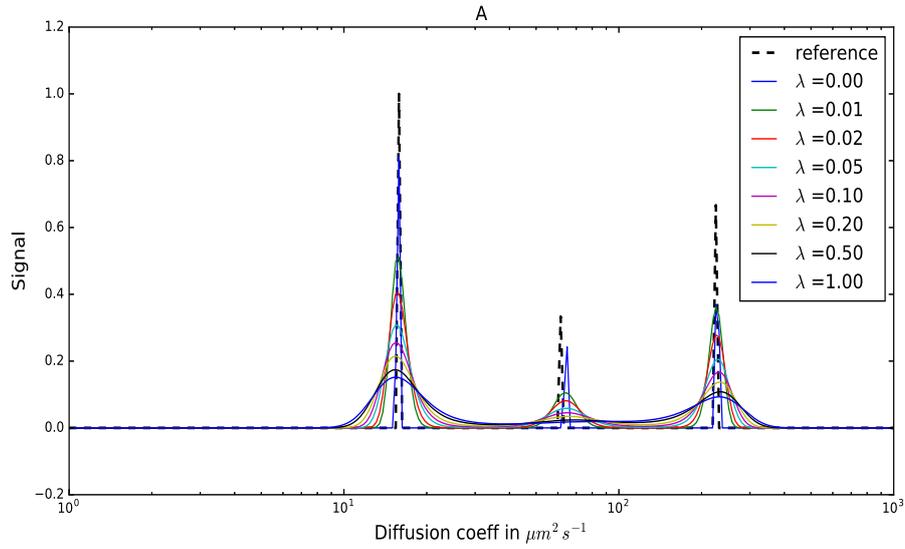

Figure S3: Reconstruction of signal A using PALMA with different $\lambda$. The minimal error is obtained for $\lambda = 0$ (Spectra obtained with $\lambda \neq 0$ have their intensities multiplied by 3 for clarity).

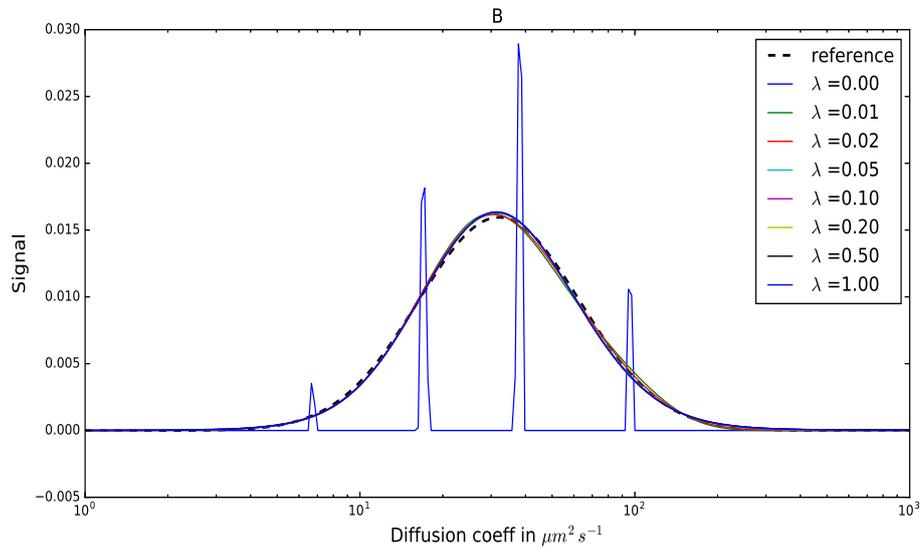

Figure S4: Reconstruction of signal B using PALMA with different $\lambda$. The minimal error is obtained for $\lambda = 0.05$ (Spectra obtained with $\lambda = 0$ have their intensities divided by 8 for clarity).

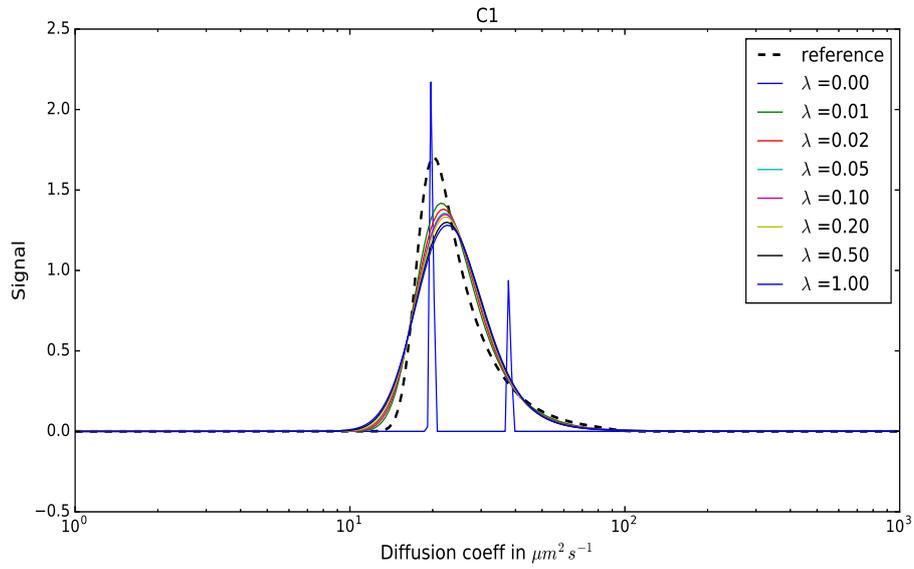

Figure S5: Reconstruction of signal C1 using PALMA with different $\lambda$. The minimal error is obtained for $\lambda = 0.01$ (Spectra obtained with $\lambda = 0$ have their intensities divided by 8 for clarity).

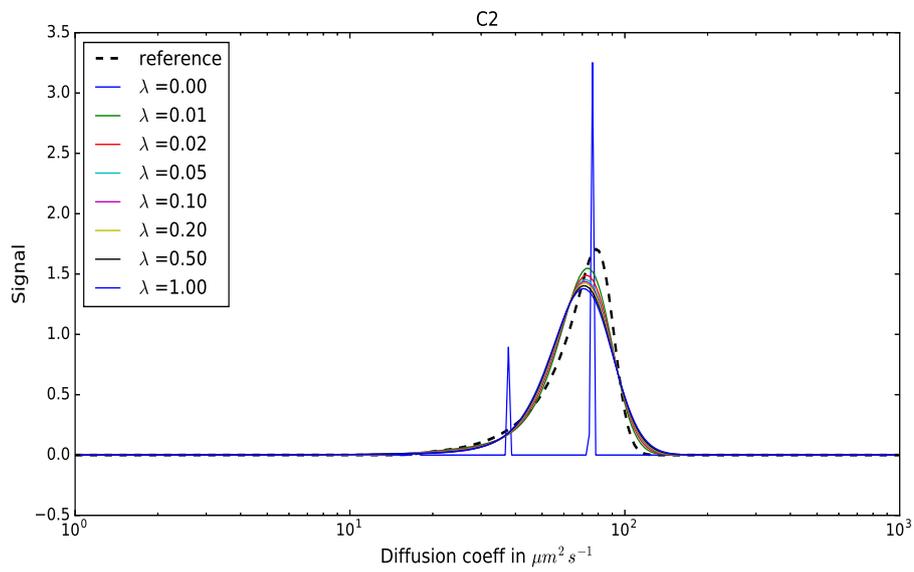

Figure S6: Reconstruction of signal C2 using PALMA with different $\lambda$. The minimal error is obtained for $\lambda = 0.01$ (Spectra obtained with $\lambda = 0$ have their intensities divided by 8 for clarity).

8

| λ | Qlty reconstruction of Signal | | | |
|---|---|---|---|---|
| | A | B | C1 | C2 |
| 0.0 | 5.65 | −11.04 | −9.87 | −12.58 |
| 0.01 | 1.05 | 28.57 | 15.21 | 16.15 |
| 0.02 | 0.80 | 31.87 | 13.76 | 14.89 |
| 0.05 | 0.58 | 32.51 | 12.92 | 14.06 |
| 0.1 | 0.47 | 31.11 | 12.71 | 13.66 |
| 0.2 | 0.39 | 31.17 | 12.37 | 13.35 |
| 0.5 | 0.30 | 31.44 | 11.65 | 12.91 |
| 1 | 0.26 | 31.55 | 11.21 | 12.44 |

Table S1: Quality of reconstruction of A, B, C1 and C2 signals for various λ values.

# 6 Comparison with state-of-the-art algorithms

Several algorithms have been developed to solve the ill-posed problem in DOSY experience. In Table S2, we present the comparison results between our approach and several recent algorithms, namely ITAMeD, ITAMeD with $\ell_p$, and TRAIN.

As an illustration, we present in Figures S7 to S10, the reconstruction of B and C2 signals with different algorithms for 4 different noise levels.

The following settings have been used, which lead to the best performance in terms of both reconstruction quality and computational cost:

PALMA: $\lambda = 0.01$.

ITAMED : Regularization parameter $= 10^{-6}$.

ITAMeD with $\ell_p$: Smoothing parameter $\tau = 10^{-7}$, ration between 1st and 2nd term $\epsilon = 10$.

TRAIn: $\tau = 1.02$

| Signal | Noise level / Algorithm | Qlty reconstruction in $dB$ | | | |
|---|---|---|---|---|---|
| | | 1% | 0.1% | 0.01% | 0.001% |
| *B* | ITAMeD | 3.37 | 18.65 | 29.04 | 29.40 |
| | ITAMeD with $\ell_p$ | 6.06 | 25.26 | 36.69 | 37.08 |
| | TRAIn | 24.75 | 28.63 | 26.53 | 19.47 |
| | PALMA with $\lambda = 0.01$ | 20.54 | 28.57 | 41.69 | 53.25 |
| | PALMA with $\lambda = 0.05$ | 24.01 | 32.51 | 48.28 | 51.37 |
| *C2* | ITAMeD | 1.57 | 15.13 | 14.36 | 14.06 |
| | ITAMeD with $\ell_p$ | 3.37 | 6.30 | 6.29 | 6.29 |
| | TRAIn | 8.62 | 15.39 | 23.36 | 25.5 |
| | PALMA with $\lambda = 0.01$ | 10.6 | 12.72 | 17.72 | 23.24 |
| | PALMA with $\lambda = 0.05$ | 7.62 | 10.97 | 16.59 | 20.75 |

Table S2: Quality of reconstruction of signals B and C2 with different algorithms for various noise levels. Here, $M = 64$

| $M$ | Noise level / Algorithm | Qlty reconstruction in $dB$ | | | |
|---|---|---|---|---|---|
| | | 1% | 0.1% | 0.01% | 0.001% |
| 64 | ITAMeD | 1.57 | 15.13 | 14.36 | 14.06 |
| | ITAMeD with $\ell_p$ | 3.37 | 6.30 | 6.29 | 6.29 |
| | TRAIn | 8.62 | **15.39** | **23.36** | **25.5** |
| | PALMA with $\lambda = 0.01$ | **10.6** | 12.72 | 17.72 | 23.24 |
| | PALMA with $\lambda = 0.05$ | 7.62 | 10.97 | 16.59 | 20.75 |
| 32 | ITAMeD | $-3.72$ | 8.89 | 12.89 | 13.91 |
| | ITAMeD with $\ell_p$ | 0.57 | 6.18 | 6.27 | 6.27 |
| | TRAIn | 3.68 | 6.60 | 9.396 | 18.04 |
| | PALMA with $\lambda = 0.01$ | **11.09** | **14.60** | **19.48** | **23.04** |
| | PALMA with $\lambda = 0.05$ | 8.69 | 13.09 | 18.92 | 20.56 |

Table S3: Quality of reconstruction of signal C2 for various noise levels and problem sizes.

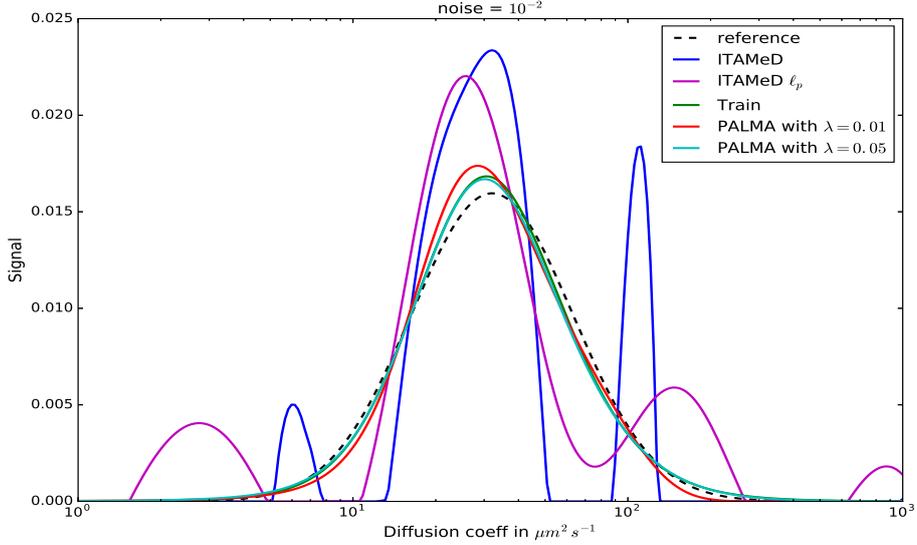

Figure S7: Reconstruction of signal B, using different algorithms. Here, $M = 64$ and noise level = 1%.

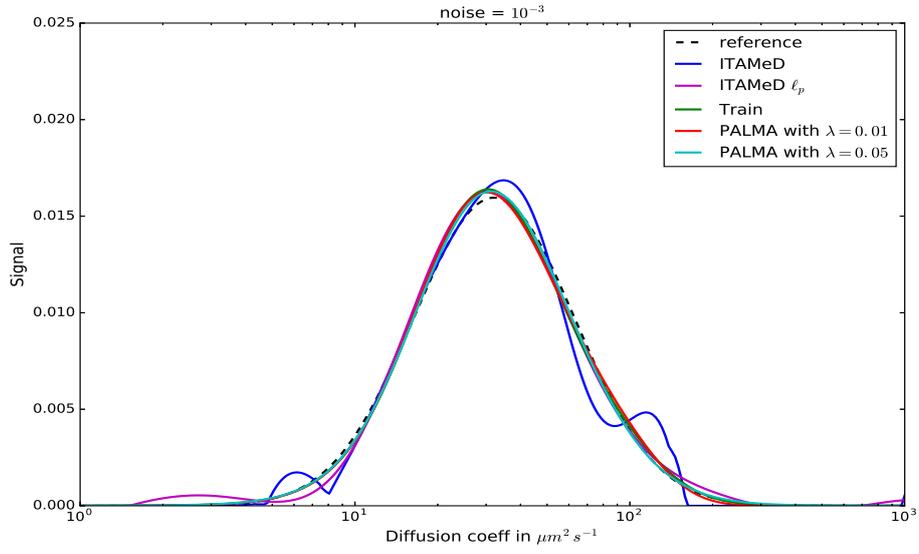

Figure S8: Reconstruction of signal B, using different algorithms. Here, $M = 64$ and noise level $= 0.1\%$.

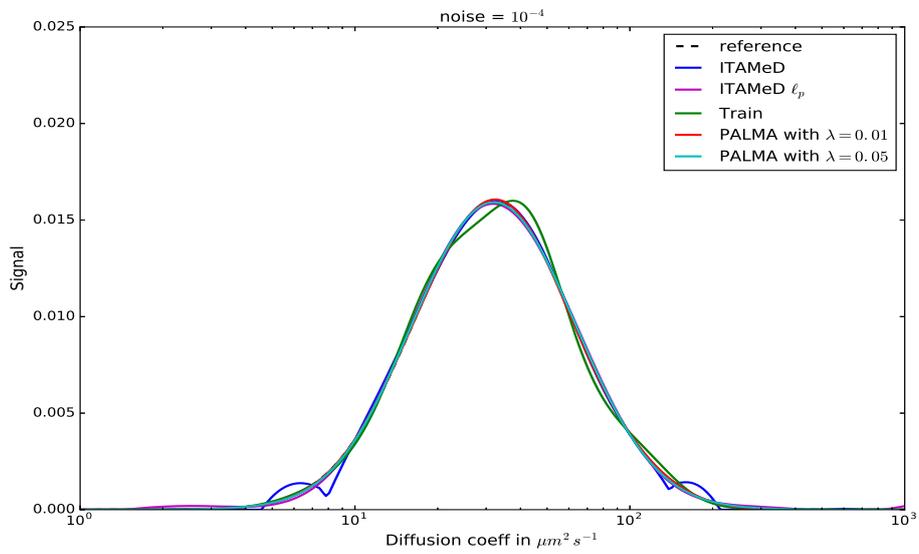

Figure S9: Reconstruction of signal B, using different algorithms. Here, $M = 64$ and noise level $= 0.01\%$.

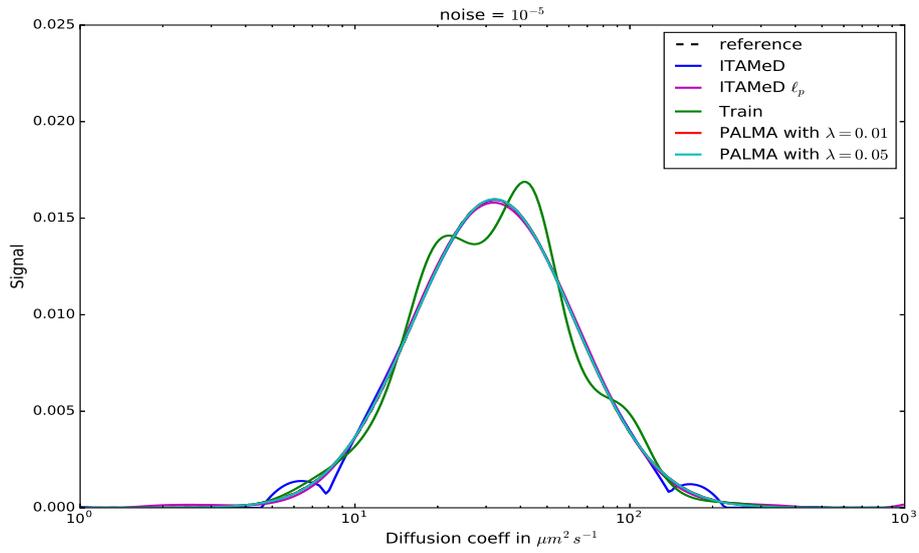

Figure S10: Reconstruction of signal B, using different algorithms. Here, $M = 64$ and noise level $= 0.001\%$.

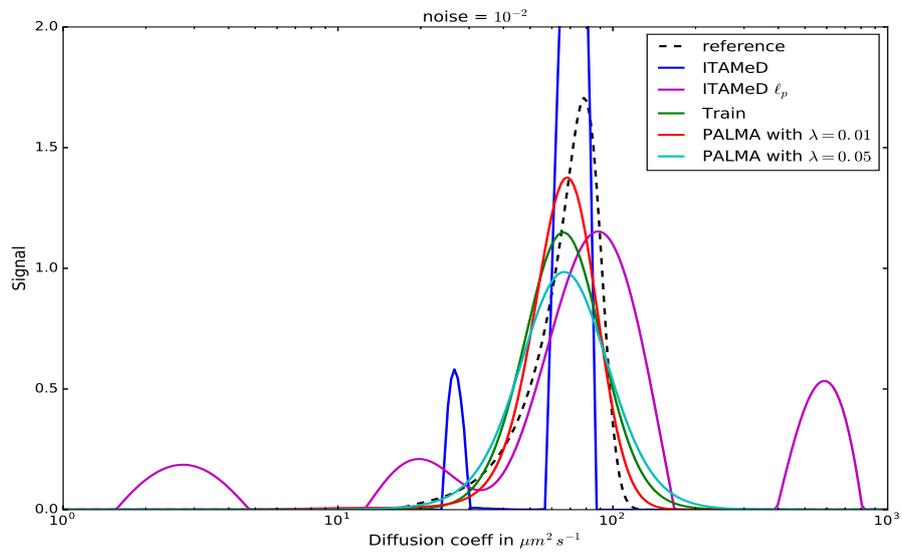

Figure S11: Reconstruction of signal C2, using different algorithms. Here, $M = 64$ and noise level $= 1\%$.

13

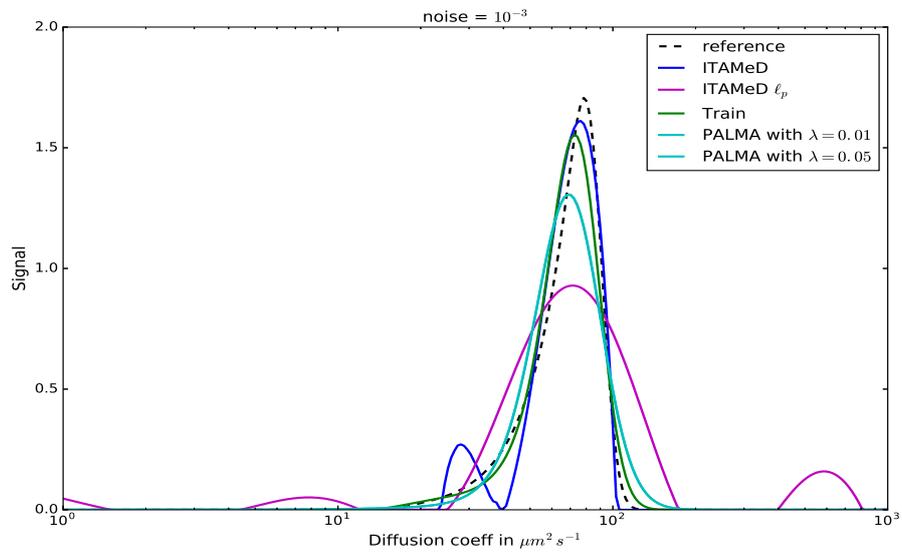

Figure S12: Reconstruction of signal C2, using different algorithms. Here, $M = 64$ and noise level = 0.1%.

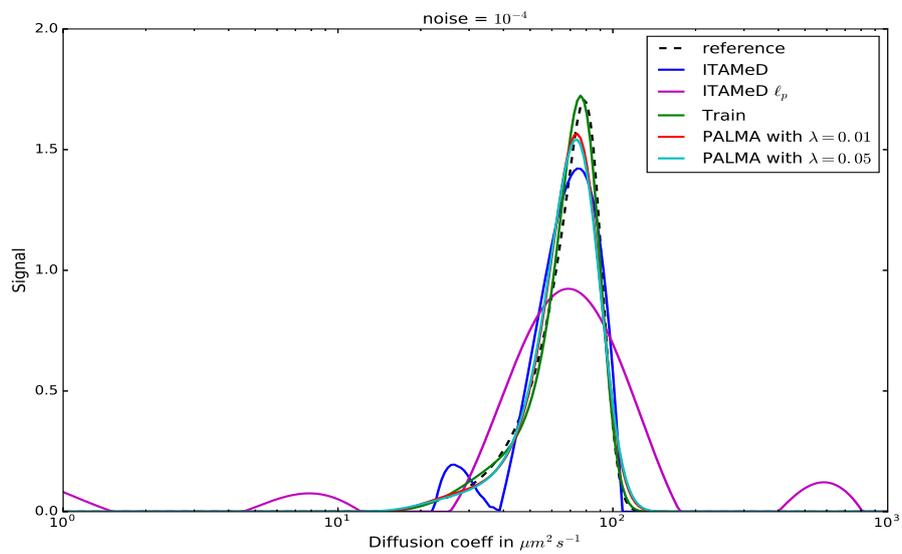

Figure S13: Reconstruction of signal C2, using different algorithms. Here, $M = 64$ and noise level = 0.01%.

14

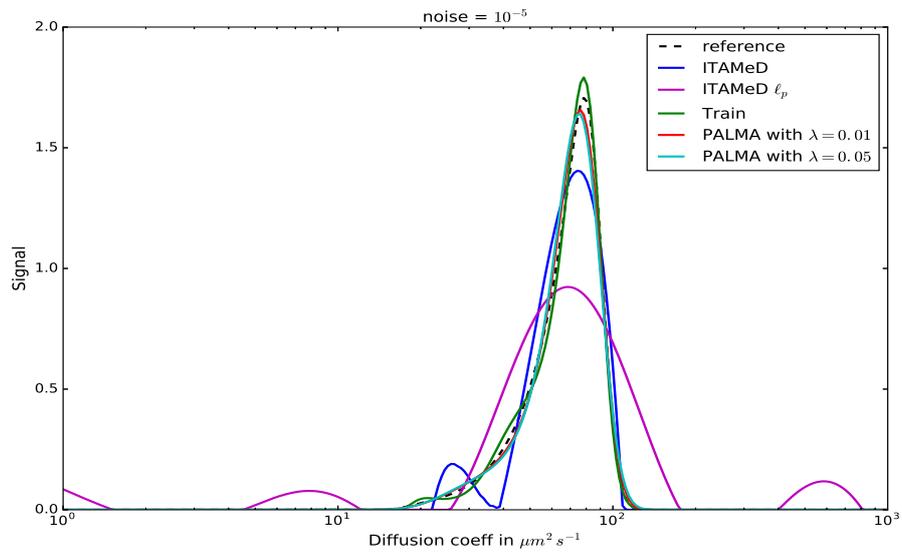

Figure S14: Reconstruction of signal C2, using different algorithms. Here, $M = 64$ and noise level $= 0.001\%$.

# 7  Robustness against noise

We propose to evaluate the quality of reconstruction of signal C2 for different $\lambda$ values and for 4 different noise levels. Figure S15 shows that the best quality reconstruction is obtained when $\lambda = 0.01$ whatever the different noise level. For this optimal $\lambda$ PALMA algorithm ensures a great quality of reconstruction as it is illustrated in Figure S16.

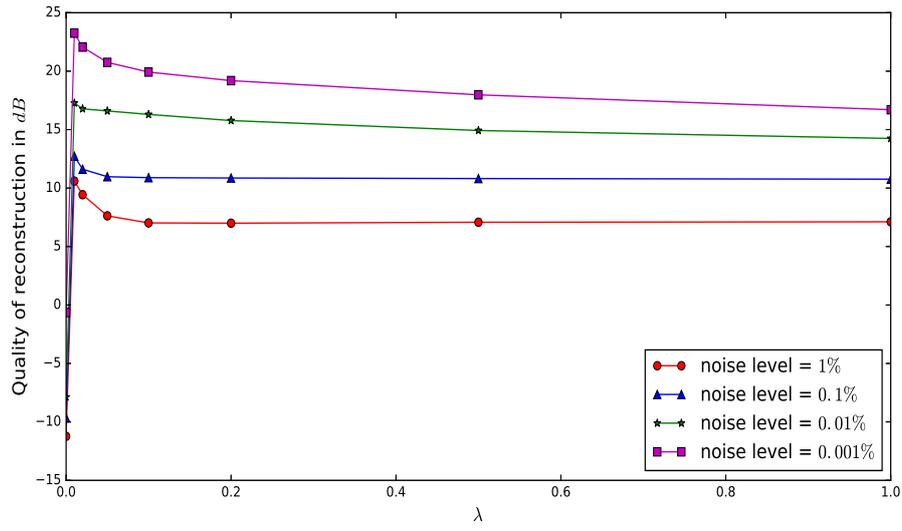

Figure S15: Quality of reconstruction of signal C2 with different $\lambda$.

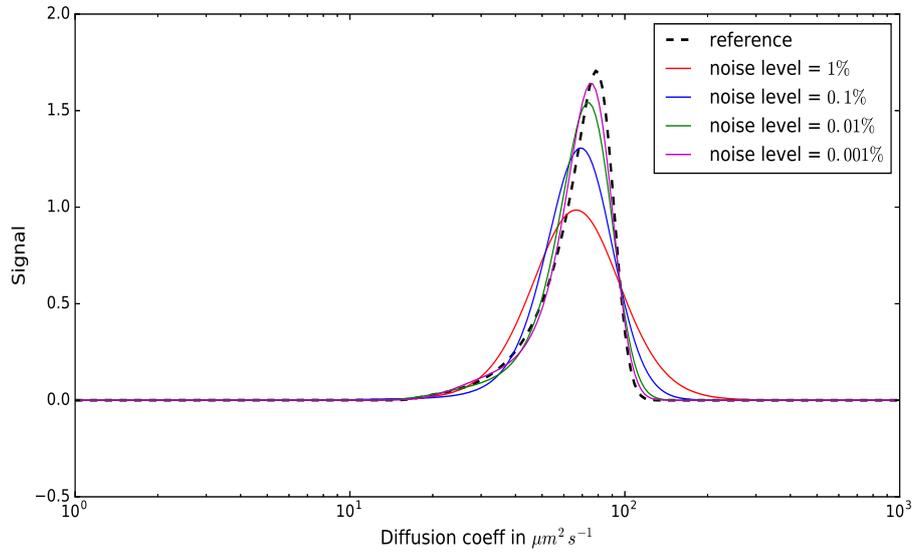

Figure S16: Reconstruction of signal C2 with $\lambda_{optimal}$ for different noise levels.

17



\end{document}